\documentclass[nofootinbib,notitlepage,onecolumn,secnumarabic,amssymb, nobibnotes, aps, prd]{revtex4-1}
\usepackage{graphicx,epsfig}
\usepackage{upgreek}
\usepackage{bbold}
\usepackage{amsmath}
\usepackage{amssymb }
\usepackage{amsfonts}
\usepackage{xcolor}
\usepackage{natbib}
\usepackage{hyperref}

\def\d {\mbox{d}}

\newcommand{\n}{{\bf {e}}}

\def\k {\bf{k}}

\def\x {\bf{x}}

\newcommand{\HI}{\text{\tiny{HI}}}
\newcommand{\HH}{\mathcal{H}}

\newcommand{\D} {\partial}

\newcommand{\<}{\langle}
\renewcommand{\>}{\rangle}

\newcommand{\three}{^{\text{\tiny (3)}}}
\newcommand{\two}{^{\text{\tiny (2)}}}
\newcommand{\one}{^{\text{\tiny (1)}}}

\def\V{^{\text{\tiny (V)}}}
\def\T{^{\text{\tiny (T)}}}

\newcommand{\p}{_{{\text{\tiny$\|$}}}}

\newcommand{\average}[1]{\left\langle #1 \right\rangle_{\tiny{\text{M}}}}

\begin{document}
\title{Nonlinear modulation of the HI power spectrum on ultra-large scales. I}
\author{Obinna Umeh$^a$, Roy Maartens$^{a,c}$ and  Mario Santos$^{a,c}$ \\~} 
\affiliation{$^a$Department of Physics, University of the Western Cape,
Robert Sobukwe Road, Bellville  7535, South Africa \\
$^b$Institute of Cosmology \& Gravitation, University of Portsmouth, Portsmouth PO1 3FX, United Kingdom\\
$^c$ SKA South Africa, The Park, Park Road, Pinelands 7405, South Africa
}

\date{\today}

\begin{abstract}
Intensity mapping of the neutral hydrogen brightness temperature promises to provide a three-dimensional view of the universe on very large scales. 
Nonlinear effects are typically thought to alter only the small-scale power, but we show how they may bias the extraction of cosmological information contained in the power spectrum on ultra-large scales. 
For linear perturbations to remain valid on large scales, we need to renormalize  perturbations at higher order. In the case of intensity mapping, the second-order contribution to clustering from weak lensing dominates the nonlinear contribution at high redshift. Renormalization modifies the  mean brightness temperature and therefore the  evolution bias. It also introduces a term that mimics white noise. 
These effects may influence forecasting analysis on ultra-large scales.

\end{abstract}

\maketitle
 \setcounter{footnote}{0}
\DeclareGraphicsRule{.wmf}{bmp}{jpg}{}{}
\maketitle

\section{Introduction}

Mapping of the integrated 21cm emission from neutral hydrogen (HI) within unresolved sources will allow us to cover very large volumes of the observable universe with upcoming radio telescopes. This promises to deliver excellent precision in measuring the standard cosmological parameters \cite{Bull:2014rha}, as well as new opportunities to probe signals on ultra-large scales, such as primordial non-Gaussianity and horizon-scale relativistic effects     \cite{Hall:2012wd,Camera:2013kpa,Santos:2015gra,Alonso:2015uua,Raccanelli:2015vla, Alonso:2015sfa, Fonseca:2015laa,Baker:2015bva}.

The ability to extract precise cosmological information from an experiment is  dependent on accurate understanding of the theoretical model. 
On scales $ k\lesssim k_{\rm nl}=0.2 h(1+z)^{2/(2+n_s)}\, \text{Mpc}^{-1}$ (where $n_s=0.96$ is the primordial spectral index), gravitational evolution of large-scale structure can be analyzed via  cosmological perturbations \cite{Smith:2002dz}. For scales $k\ll k_{\rm nl}$,
 it is commonly assumed that one need not go beyond linear order. This is valid for cold dark matter (CDM) perturbations, which obey precise energy and momentum conservation, but not necessarily for tracers of the underlying CDM distribution.

 It was shown in \cite{Heavens:1998es} that mode coupling in the  biased density field contributes power on large scales which  becomes important on Hubble scales.
This potential threat to the validity of linear perturbation theory on large scales can be mitigated by renormalization  \cite{McDonald:2006mx,McDonald:2009dh,Chan:2012jx,Nishizawa:2012db, Assassi:2014fva}.
Renormalization  leads to a change in the mean (zero-order) density and in the bias parameters for the tracer. In addition, it introduces a noise-like term \cite{McDonald:2006mx,McDonald:2009dh,Assassi:2014fva}.

We employ this renormalization technique to study the HI brightness temperature after reionization. Up to second order in perturbation theory, the mean HI brightness temperature is modified by a term that depends on redshift and a coarse-graining scale. At $z \lesssim 1.5$, this term is proportional to a nonlinear bias parameter, while
for $z \gtrsim 1.5$, it is dominated by a nonlinear weak lensing term. 
We point out that the renormalization of mean brightness temperature leads to a renormalization of the evolution bias parameter, which modifies the linear power spectrum on horizon scales.
The main implication is that cosmological constraints on horizon scales need to incorporate nonlinear effects that flow from renormalization.

Our fiducial model is determined by the Planck 2015 best-fit values \cite{Ade:2015xua}; in particular,  $h =H_0/100= 0.673$,   $\Omega_{m0} =1-\Omega_{\Lambda 0}= 0.315$. A busy reader can skip section \ref{sec:perturbation} which covers the perturbation theory (with details given in Appendix A). Section \ref{sec:renormalization} covers the  nonlinear (Gaussian) bias (with details for HI in Appendix B), and the basic ideas of renormalization are introduced and then implemented in redshift space. We discuss our results in section \ref{sec:discus} and conclude in section \ref{sec:conc}. 

\section{Fluctuations in HI brightness temperature}\label{sec:perturbation}

The brightness temperature observed (in redshift space) is given by \cite{Hall:2012wd}
\begin{eqnarray}\label{eq:arbbrightness}
T^{\rm obs}(z,{\n})&=&\frac{3\pi^2}{4} \frac{\hbar^3 A_{10} }{k_{B}E_{21}}\,n(z,{\bf e})J(z,{\bf e})\,,
\end{eqnarray}
 where $z $ is the redshift  of the source, ${\n}$ is the unit direction of the source,    $E_{21}$ is the proper energy of the emitted photons, and $A_{10}= 2.869 \times 10^{-15} {\rm s}^{-1}$ is the emission rate.  The number density of HI atoms is $n$ and $J$ is the Jacobian of the map between real and redshift space [see  \eqref{eq:dhatlamdbyz}]. 

The observed temperature is expanded up to second order: 
    \begin{eqnarray}\label{eq:brightHI}
T^{\rm obs}(z,{\n})&=&  \bar{T} (z)\Big[1+\Delta_{T}\one(z,{\n})+\frac{1}{2}\Delta\two_{T}(z,{\n})\Big]\,,
\end{eqnarray}
where we use $\Delta_T$  to denote the  quantity observed in redshift space. The mean brightness temperature is 
  \begin{eqnarray}\label{eq:backgdeltaTbin}
  \bar{T} (z) &=&\frac{3\pi^2}{4} \frac{\hbar^3 A_{10}}{k_{B}E_{21}}\,
\frac{ \bar{n} (z)a(z)^3}{\HH(z)}\approx  566 h \frac{\Omega_{\HI}(z)}{0.003}(1+z)^2\frac{H_0}{\HH(z)} \,\,\mu {\rm K}.
\end{eqnarray}
Here $\Omega_{\HI}$ is the comoving HI mass density
in units of the current critical density. 

The observed fractional temperature perturbation $\Delta_T$ is related to the fractional number overdensity $\delta_n$ in real space as follows (see Appendix A for details):
  \begin{eqnarray}
\label{eq:firstdeltaTb1in}
\Delta_{T}\one(z,{\n})&=&\delta_n  \one+\delta\one z-\delta\one k^0 +\frac{1}{\HH} \frac{\d\delta\one z}{\d\lambda}+\left(b_e-2\HH -\frac{\HH'}{\HH}\right)\delta\one \lambda\,,\\
\Delta\two_{T}(z,{\n})&=&
\delta\two_n +\delta\two z-\delta\two k^0-\omega_{\p}\two+\frac{1}{\HH}\frac{\d\delta\two z}{\d\lambda}  + \left(b_e-2\HH-\frac{\HH'}{\HH}\right)\delta\two \lambda \nonumber\\&&
+ 2\left[\delta\one k^0 - \delta\one z- \frac{1}{\HH}\frac{\d\delta\one z}{\d\lambda}\right]\left[\delta\one k^0 -\delta_n\one  \right]
+2\left[\frac{1}{\HH}\frac{\d\delta\one z}{\d\lambda}\right]^2\nonumber \\ \nonumber&&
+2\delta\one\lambda\left[\left(\delta_n\one  -\frac{1}{
\HH}\frac{\d\delta\one z}{\d\lambda}+\delta\one z-\delta\one k^0\right)
\left(b_e-2\HH-\frac{\HH'}{\HH}\right)-\frac{\HH'}{\HH^2}\frac{\d\delta\one z}{\d\lambda}\right]
\\ \nonumber&&
+[\delta\one\lambda]^2 \left[b_e^2 +\frac{\d b_e}{\d\lambda}-4\HH\left(b_e-\HH\right)-2\frac{\HH'}{\HH}\left(b_e-2\HH-2\frac{\HH'}{\HH}\right)-\frac{\HH''}{\HH}\right]
\\&&
+2\left[\delta\one \lambda {\Delta_{T}\one}' + \Delta x_{\p}\one \partial_{\p} \Delta_{T}\one + \Delta x_{\bot}^{\text{\tiny (1)} i}\nabla_{\bot i} \Delta_{ T}\one\right]\,. \label{eq:seconddeltaTb1in}
\end{eqnarray}
Here 
$ k^a=\d x^a/\d\lambda$ is the photon 4-momentum, $\Delta x^a$ is the perturbed physical position, $\omega$ is the metric vector perturbation and $\|$ and $\bot$ denote projection along and transverse to $\n$. 
The evolution bias is defined by the mean number density: 
 \begin{equation}\label{eq:evolbias}
     b_e= -\HH(z)\frac{\d\ln [\bar{n} (1+z)^{-3}]}{\d\ln(1+ z)}.
     \end{equation}
From now on, for convenience we will often drop the superscript (1) for first-order perturbations.

\begin{itemize}

\item {{\bf Background:}} We calculate $\Omega_{\HI}$ from a simple halo model  using
\begin{equation}\label{eq:OMHI}
\Omega_{\HI}(z)=\frac{1}{(1+z)^3}\frac{\rho_{\HI}(z)}{\rho_{c0}}\,,\quad \mbox{with}\quad
\rho_{\HI}(z) = \int_{M_{\text{min}}}^{M_{\text{max}}}\d M \, n_h(M) M_{\HI}(M)\,,
\end{equation}
where $\rho_{c0}= 3 H^2_0/(8 \pi G)$ and $n_h$ is the halo mass function. See Appendix B for details.

\item {{\bf First-order perturbation:}}
In Poisson gauge [see  (\ref{eq:metric})], the full first-order perturbation  is given by \eqref{eq:HIoverdensity}:   
     \begin{eqnarray}\label{eq:HIoverdensitymain}
      \Delta_{T}\one(z,{\n})&=&\delta_n  \one-\frac{1}{\HH}\partial^2_{\p}v+
      \frac{1}{\HH}\left(b_e-2\HH-\frac{\HH'}{\HH}\right)\partial_{\p}v+\frac{1}{\HH}\Psi'-\frac{1}{\HH}\left(b_e-3\HH-\frac{\HH'}{\HH}\right)
     \Phi
     \\ \nonumber&&
     +\frac{1}{\HH}\left(b_e-2\HH-\frac{\HH'}{\HH}\right)
\int_{\lambda_o}^{\lambda}\d \tilde\lambda (\Phi' + \Psi')\,, 
     \end{eqnarray}
where $v$ is the velocity potential.
We have dropped the subscript $s$ for the source and set the observer-dependent monopole and dipole part to zero [see \eqref{eq:HIoverdensity}]. Equation (\ref{eq:HIoverdensitymain}) is in agreement with  (17) of \cite{Hall:2012wd}.

The second term is the linear Kaiser redshift space distortion (RSD) term, followed by the Doppler term, the rate of change of the curvature potential, the gravitational redshift term and finally the integrated Sachs-Wolfe term.
      The third to the last terms are due to relativistic lightcone effects, and are suppressed on subhorizon scales \cite{Hall:2012wd}.  Note that there is no weak lensing contribution  to the HI temperature at first order.
      
      A further horizon-scale term arises when we introduce the scale-independent bias.  The relation $\delta_n=b\delta_m$, where $b=b(z)$, is only physical (and thus gauge-invariant) if expressed in the matter rest-frame, i.e. in comoving-synchronous gauge  \cite{Hall:2012wd}:
\begin{equation}\label{bias}
\delta\one_n=b\delta^{{\rm cs}(1)}_m+(3\HH-b_e) v.
\end{equation}
     
     \item {{\bf Second-order perturbation:}}
The second-order perturbations of the galaxy number counts that are observed on the lightcone in redshift space, were recently computed by
\cite{Bertacca:2014dra,Bertacca:2014wga,Yoo:2014sfa,DiDio:2014lka,Bertacca:2014hwa}. The extremely lengthy expressions are however not directly applicable to the case of intensity mapping which we are treating. 
In Appendix A we compute $ \Delta\two_{T}(z,{\n})$, using the approach developed in \cite{Umeh:2012pn,Umeh:2014ana}. The full expression is unmanageable, and we make a consistent approximation that generalizes the  Kaiser approximation \cite{Kaiser:1987qv} to second order:
 \begin{eqnarray}\label{eq:SimpKaiser1}
           \Delta\two_{T}(z,{\n})&= &\delta_n  \two-\frac{1}{\HH}\partial^2_{\p}v\two \overbrace{-\frac{2}{\HH}\delta_n  \partial^2_{\p}v+2\left(\frac{1}{\HH}\partial^2_{\p}v\right)^2
           -\frac{2}{\HH}\partial_{\p}v\left[\partial\p \delta_n  -\frac{1}{\HH}\partial\p^3 v \right]}^{\text{nonlinear RSD}}
            \\ \nonumber&&
            +\underbrace{2\left[\nabla_{\bot i} \delta_n  -\frac{1}{\HH}\nabla_{\bot i}\partial\p^2 v \right]\int_{0}^{\chi}\d\tilde \chi(\tilde\chi-\chi) (\nabla^i_{\bot}\Phi+ \nabla_{\bot}^i \Psi)}_{\text{nonlinear lensing}}+\underbrace{\frac{2}{\HH}\partial\p v \left[\delta'_n  -\frac{1}{\HH_s}\partial\p^2v'\right]}_{\text{nonlinear time perturbation}}.
       \end{eqnarray}
The nonlinear RSD terms        
 recover the result of \cite{Heavens:1998es,Verde:1998zr} for the galaxy number density. 
The nonlinear lensing term is $ \Delta x_{\bot}^{\text{\tiny (1)} i} \nabla_{\bot i} \Delta_{ T}\one$, i.e. the contraction of the weak lensing deflection angle with the transverse gradient of the first-order (density + RSD) term. This is the dominant nonlinear correction to \cite{Heavens:1998es,Verde:1998zr}.
The time perturbation term is subdominant in the power spectrum -- but it may be more important in the bispectrum. There are many further terms  that we have neglected because they are strongly suppressed on sub-Hubble scales.
  
\end{itemize}

\subsection*{Relationship with galaxy number density}

  We derive a covariant  mapping between the perturbation of observed brightness temperature and the perturbation of observed galaxy number density. 
 If we count the observed number of HI atoms per solid angle and per redshift instead of galaxies, we find that the observed brightness temperature is given by \cite{Hall:2012wd}
   \begin{eqnarray}\label{eq:brightness}
T^{\rm obs}(z,\n)&=&\frac{3\pi^2}{4} \frac{\hbar^3 A_{10}}{k_{B}E_{21}^2}\frac{n^{\tiny{\text{obs}}}(z,\n)}{D_{A }(z)^2}\,,
\end{eqnarray}
where  $D_A$ is the angular diameter distance.
Expanding in perturbations we find
 \begin{eqnarray}\label{eq:compare1}
 \Delta_{T}\one &=& \Delta_n\one - 2\frac{\delta  D_A}{\bar{D}_A}\,,\\
 \Delta^{(2)}_{T}&=& \Delta^{(2)}_n - 2\frac{\delta\two  D_A}{\bar{D}_A}-4\Delta_n\frac{\delta  D_A}{\bar{D}_A}+6\left(\frac{\delta  D_A}{\bar{D}_A}\right)^2-4\left[\delta \lambda \left(\frac{\delta D_A}{\bar{D}_A}\right)' + \Delta x_{\p} \partial_{\p}  \left(\frac{\delta D_A}{\bar{D}_A}\right) + \Delta x_{\bot}^i\nabla_{\bot i} \left(\frac{\delta D_A}{\bar{D}_A}\right)\right].
 \label{eq:compare2}
 \end{eqnarray}
 Here $\Delta_n$ is the fractional HI galaxy number overdensity observed in redshift space, before magnification bias is applied.
 At linear order, there is an accidental symmetry whereby $\Delta\one_{T}$ is obtained from the magnification bias corrected $\Delta_n$, by simply setting the slope of the luminosity function to $s=2/5$ \cite{Hall:2012wd}.
  Equation (\ref{eq:compare1}) is in exact agreement with  (23) of \cite{Hall:2012wd} and  (\ref{eq:compare2}) is its second-order extension.

\section{Nonlinear perturbations and renormalization}\label{sec:renormalization}

In order to illustrate the main idea, we start by
ignoring the corrections due to observation in redshift space and by focusing on sub-Hubble scales, where the difference between comoving-synchronous and Newtonian gauges in  \eqref{bias} may be ignored.
Fluctuations in number density should average to zero over large enough volumes, i.e. 
\begin{equation}\label{eq:def}
\delta_n  (z,{\x}) = \frac{n (z,{\x}) -\<{n}  \>(z) }{\<{n}  \>(z) } \quad \Rightarrow \quad \<\delta_n  (z,{\x})\> = 0.
\end{equation}
 Within standard cosmological perturbation theory, we normally assume that $\<{n}  \>(z) = \bar{n} (z) = \bar{n} ^{\text{FLRW}}(z)$, i.e we assume that $\bar{n} $ is determined by the background FLRW spacetime. However this does not take account of the complexities of relating a tracer to the underlying CDM distribution.

A bias model is needed  to relate $\delta_n  $ to the CDM overdensity $\delta_m$. We assume that the primordial perturbations are Gaussian, and use an Eulerian local bias model, applied up to second order:
\begin{eqnarray}\label{eq:biasillust}
\delta_n(z,{\x})  = b_1(z)   \delta_m(z,{\x}) +\frac{1}{2} b_2(z)   \big[\delta_m (z,{\x})\big]^2\,.
     \end{eqnarray}
As a consequence, the average in  (\ref{eq:biasillust}) no longer vanishes:
\begin{equation}\label{reno}
 \< \delta_n  \>(z)= b_2(z)\sigma^2_S(z),\quad \sigma^2_{S}(z) = \int_{k_{\text{min}}}^{k_S} {\d^3 k \over (2\pi)^3} P_{m}(z,k).
\end{equation}
Here 
$\sigma^2_{S}$  is the variance of $\delta_m$ and  $k_S$ is the small-scale cutoff. 

A mathematically consistent way of ensuring zero average  is via renormalization \cite{McDonald:2006mx,McDonald:2009dh,Assassi:2014fva,Nishizawa:2012db}.  
This involves subtracting $b_2\sigma^2_S/2 $ from both sides of  (\ref{eq:biasillust}).  The right-hand side will then have zero average, by  \eqref{reno},  while the left-hand side will define a renormalized fluctuation $\delta^R_n=\delta_n-b_2\sigma^2_S/2$. 

Renormalization of $\delta_n$ must leave invariant the physical number density $n$. This is achieved by renormalization of the background number density, $\bar n$, which in turn induces a renormalization of the bias parameters:
\begin{equation}\label{reno2}
\bar{n}^R = \bar{n} \big(1 + b_2   \sigma^2_{S}/2\big),~~~b_a^R={b_a \over \big(1 + b_2   \sigma^2_S/2\big)}.
\end{equation}
Putting this together, we have
\begin{eqnarray}\label{eq:renormalizedn}
 \delta_n  ^{{R}} \equiv \frac{n  -\bar{n} ^R}{\bar{n} ^R}=   b_1^R\delta_{m} + 
\frac{1}{2}{b_2^{R}}\left[ (\delta_m)^2 - \sigma^2_{S}\right]
  \quad \Rightarrow \quad \<\delta_n  ^{{R}}\> =0.
\end{eqnarray}
Note that the fundamental physical quantity (i.e $n $) is independent of the cutoff scale chosen to regulate the integral in $\sigma^2_{S}$, i.e.,  $\partial n /\partial k_S = 0$ \cite{Assassi:2014fva}. 

An important consequence of renormalization  was not mentioned in \cite{McDonald:2009dh,Assassi:2014fva,Nishizawa:2012db}, since these papers did not include the horizon-scale  general relativistic  terms that we have in  \eqref{eq:HIoverdensitymain}. This consequence is that the evolution bias $b_e$,  which appears in the coefficients of most of the general relativistic terms, will be renormalized as a result of renormalization of the mean number density.  Using equations \eqref{eq:evolbias}  and \eqref{reno2}, $b_e$ is modified by $n$-renormalization as
\begin{equation}
 b_e^R   = b_e - \HH(z)\frac{\d \ln(1+b_2   \sigma^2_{S}/2)}{ \d \ln (1+z)}\,.
 \label{eq:evolbias2}
 \end{equation}
As a result, the linear power spectrum of the observed fractional temperature fluctuations will be modified on horizon scales by renormalization. 
 
The power spectrum may be computed by expanding  (\ref{eq:renormalizedn}) in Fourier space (we suppress the $z$-dependence):
\begin{eqnarray}
\delta_n  ^{R}({\k})=b_{1}^{R}\delta_m({\k})
+ \frac{1}{2}\bigg[ \int\frac{\d^3 k_1}{(2\pi)^3}\frac{\d^3 k_2}{(2\pi)^3}\delta_{m}({\k}_1)\delta_{m}({\k}_2)(2\pi)^3\delta^{D}\!({\k}-{\k}_1-{\k}_2)\left[b_{2}^{R} + b_{1}^{R} F_2({\k}_1,{\k}_2)\right]
- b_{2}^{R}\sigma^2_{S}\,\delta^D\!({\k}) \bigg],
\end{eqnarray}
where $P_m$ is the linear matter power spectrum and $F_2$ is the standard nonlinear density kernel, which we approximate using a matter-dominated model:
\begin{eqnarray}\label{eq:F2}
F_2({\k}_1,{\k}_2)=\frac{5}{7}+\frac{1}{2} \left({k_1 \over k_2}+{k_2 \over k_1}\right) {{\k}_1 \cdot {\k}_2 \over k_1
k_2} +\frac{2}{7}
\left({{\k}_1 \cdot {\k}_2 \over k_1
k_2}\right)^2.
\end{eqnarray}

Computing $\< \delta_n^R({\k})\delta_n^R({\k}')\>$, we find that the contributions of the $\sigma_S^2$ term exactly cancel out and the power spectrum of the renormalized fractional number overdensity is
\begin{eqnarray}\label{unren}
{P}_{\delta^R_n}(k) =\left(b_{1}^{R}\right)^2P_{m}(k)
+\frac{1}{2}\int \frac{\d^3k_1}{(2\pi)^3}\Big[b_{2}^{R} + b_{1}^{R} F_2({\k}_1,{\k}-{\k}_1)\Big]^2P_{m}(|{\k-{\k}_1|})P_{m}({k}_1) .
\end{eqnarray}
The  $b_2^R$ term leads to constant power on very large scales \cite{Heavens:1998es,McDonald:2009dh,BeltranJimenez:2010bb}:
\begin{equation}
k\to 0 ~\Rightarrow~ {P}_{\delta^R_n}(k)  \to \frac{1}{2}\left(b_{2}^{R}\right)^2  \int \frac{\d^3k_1}{(2\pi)^3}P_{m}^2({k}_1),
\end{equation}
where we used $F_2({\k},-{\k})=0$. It is possible to renormalize the power spectrum by subtracting this constant power from both sides of  \eqref{unren}. In this approach, the constant-power contribution is treated as part of the noise budget \cite{Heavens:1998es,McDonald:2006mx,McDonald:2009dh}:
\begin{eqnarray}\label{eq:powernumber}
{P}^R_{\delta^R_n}(k)&=&\left(b_{1}^{R}\right)^2P_{m}(k)
+\frac{1}{2}\int \frac{\d^3k_1}{(2\pi)^3}\left\{\big[b_{2}^{R} + b_{1}^{R} F_2({\k}_1,|{\k-{\k}_1|})\big]^2P_{m}(|{\k-{\k}_1|})-\left(b_{2}^{R}\right)^2P_m(k_1)\right\}P_{m}({k}_1) ,
\\ {P}_{\delta^R_n}(k)&=&{P}^R_{\delta^R_n}(k)+N_{\rm eff},~~ N_{\rm eff}=\frac{1}{2}\left(b_{2}^{R}\right)^2  \int \frac{\d^3k_1}{(2\pi)^3}P_{m}^2({k}_1).
\end{eqnarray}

 Here we follow a different approach, consistent with \cite{Jeong:2008rj,BeltranJimenez:2010bb}. We see the constant-power term as containing cosmological information rather than noise, i.e. we treat it as a nonlinear correction to the power spectrum. In this view, we use the unrenormalized power spectrum \eqref{unren}. 
The two approaches differ only in  the interpretation of the correction to the linear power spectrum on ultra-large scales.

\subsection*{Renormalization in redshift space for the brightness temperature}\label{sec:rsd}

We now extend the established renormalization procedure from real space to the observed redshift space, in the case of the observed fractional HI  brightness temperature perturbation:
\begin{eqnarray}\label{eq:brightHI}
\Delta_{T}(z,\n ) = \frac{T^{\text{obs}}(z,\n ) - \<T^{\text{obs}}\>(z)}{\<T^{\text{obs}}\>(z)}.
\end{eqnarray}
The sky average of $\<\Delta_{T}\>$ is zero by definition, but if we assume that $\<T^{\text{obs}}\>= \bar{T} ^{\text{FLRW}}$, we get a nonzero value. In order to ensure that $\<\Delta_{T}\> = 0$, we have to renormalize the background temperature.  

In observed redshift space, the average is more complicated because of the nonlinear lensing contribution  in  \eqref{eq:SimpKaiser1}. After performing the sky integral, we find that
\begin{eqnarray}\label{eq:expectionvalue}
\<\Delta_{T}\>(z) =  \frac{1}{2}\<\Delta_{T}\two\>(z)
= \left\{  b _2(z) + 4\Big[ b _1(z) + \frac{1}{5} f(z)\Big]T_{\kappa}(z)\right\}\sigma^2_{S}(z)\,,
\end{eqnarray}
where $f=\d\ln D/\d\ln a$ is the growth rate and $D$ the growing mode of the linear matter overdensity, and
\begin{equation}
T_{\kappa}(z)
=\frac{\Omega_{m0}H_0^2}{D(z)}\int_{0}^{\chi}\d\tilde\chi \,[1+z(\tilde\chi)] D(\tilde\chi)(\tilde\chi-\chi){\tilde\chi \over \chi },
\end{equation}
is a transfer function for the weak lensing convergence.
The sky average of the nonlinear lensing term is the only redshift-space contribution to the monopole.

We renormalize the background temperature and its fluctuations by adding and subtracting $\<\Delta_{T}\>$ to  (\ref{eq:brightHI}): 
\begin{eqnarray}
\bar{T} ^R &=& \bar{T}\big(1+  \<\Delta_{T}\>\big),\\
\label{eq:TbFourier}
\Delta_{T}^R({\k},\mu) &=& {\Delta_{T}^R}\one({\k},\mu)+\frac{1}{2}\left[{\Delta_{T}^R}\two({\k},\mu)-\<\Delta_{T}\>\delta^{D}({\k})\right] ,\quad \mu={{\k}\cdot {\n}\over k},~~ \<\Delta_{T}^R({\k},\mu)\> =0,
\end{eqnarray}
where we suppressed the $z$-dependence. Here the renormalized first- and second-order parts are
 \begin{eqnarray}\label{eq:z1}
{\Delta_{T}^R}\one({\k},\mu) &=& { \delta_m({\k}) \over 1+\<\Delta_{T}\>} \left[b _1 +f\mu^2+\mathcal{A}\,\frac{\HH^2}{k^2}+i \mu\mathcal{B}\, \frac{  \HH}{k}\right] \,,\\
{\Delta_{T}^R}\two({\k},\mu) &=&{1 \over 1+\<\Delta_{T}\>}  \int\frac{\d^3 k_1}{(2\pi)^3}\frac{\d^3 k_2}{(2\pi)^3}\delta_{m}({\k}_1)\delta_{m}({\k}_2)(2\pi)^3\delta^D\!({\k}-{\k}_1-{\k}_2)\bigg[ b_2+b_1    F_2({\k}_1,{\k}_2) 
 \nonumber\\ &&
+ f \mu^2 G_2 ({\k}_1,{\k}_2)  + \mathcal{K}_{\text{R}}({\k}_1,{\k}_2)
+\mathcal{K}_{\text{L}}({\k}_1,{\k}_2)
+i\mathcal{K}_{\text{T}}({\k}_1,{\k}_2)\bigg].
\label{eq:z2}
\end{eqnarray}
In order to define the HI bias physically at linear order, we used the  definition \eqref{bias} and we applied \eqref{eq:biasillust} in comoving-synchronous gauge. For convenience we have omitted the cs superscript -- i.e., $\delta_m$ in the above and following equations is understood to be $\delta^{\rm cs}_m$. Note that we do not need to extend the comoving-synchronous definition to second order, because the difference contains only terms that we are neglecting at second order.

The $\mathcal{A}$ and $\mathcal{B}$ terms in  (\ref{eq:z1}) are the coefficients of the first-order potential and Doppler terms in \eqref{eq:HIoverdensitymain}: 
     \begin{eqnarray}
     \mathcal{A}& = &f\left(3-\frac{b_e^R}{\HH} - \frac{3}{2}\Omega_m\right) 
     - \left[2-\frac{b_e^R}{\HH}+\frac{\HH'}{\HH^2}\right]
     \left[ \frac{3}{2}\Omega_m +3\int_0^{z} \d \tilde z
     \,\Omega_m(\tilde z) \frac{\HH^2(\tilde z)D(\tilde z) }{\HH^2( z) D(z)}[f(\tilde z)-1]
     \right],\\
     \label{eq:kernelA}
     \mathcal{B}& =& -f\left(2-\frac{b_e^R}{\HH}+\frac{\HH'}{\HH^2}\right)\,.
     \label{eq:kernelB}
     \end{eqnarray}
The renormalized background temperature is equivalent to a renormalized background number density by \eqref{eq:backgdeltaTbin}, so that the evolution bias  is renormalized:  
\begin{eqnarray}\label{renormevo}
      b_e^R   = b_e - \HH(z) \frac{\d\ln \big(1+\<\Delta_{T}\>_L\big)}{ \d \ln (1+z)}\,.
\end{eqnarray}
We have introduced a smoothing scale $L$ (in units of $ {h}^{-1}\,$Mpc). For a top-hat window function on a scale $L$,  (\ref{eq:expectionvalue}) becomes:
\begin{equation}
\<\Delta_{T}\>_{L} =  \int \frac{\d^3 k}{(2\pi)^3} P_m (k)\left[ \big|W({kL})\big|^2 b _2 + 4 W({kL})\left(  b _1 + \frac{1}{5} f\right)T_{\kappa}\right]\,, \quad W({kL}) = 3 \left[\frac{\sin(kL)}{(kL)^3} -\frac{\cos(kL)}{(kL)^2}\right].
\end{equation}

\begin{figure}[htb!]
\includegraphics[width=0.7\textwidth]{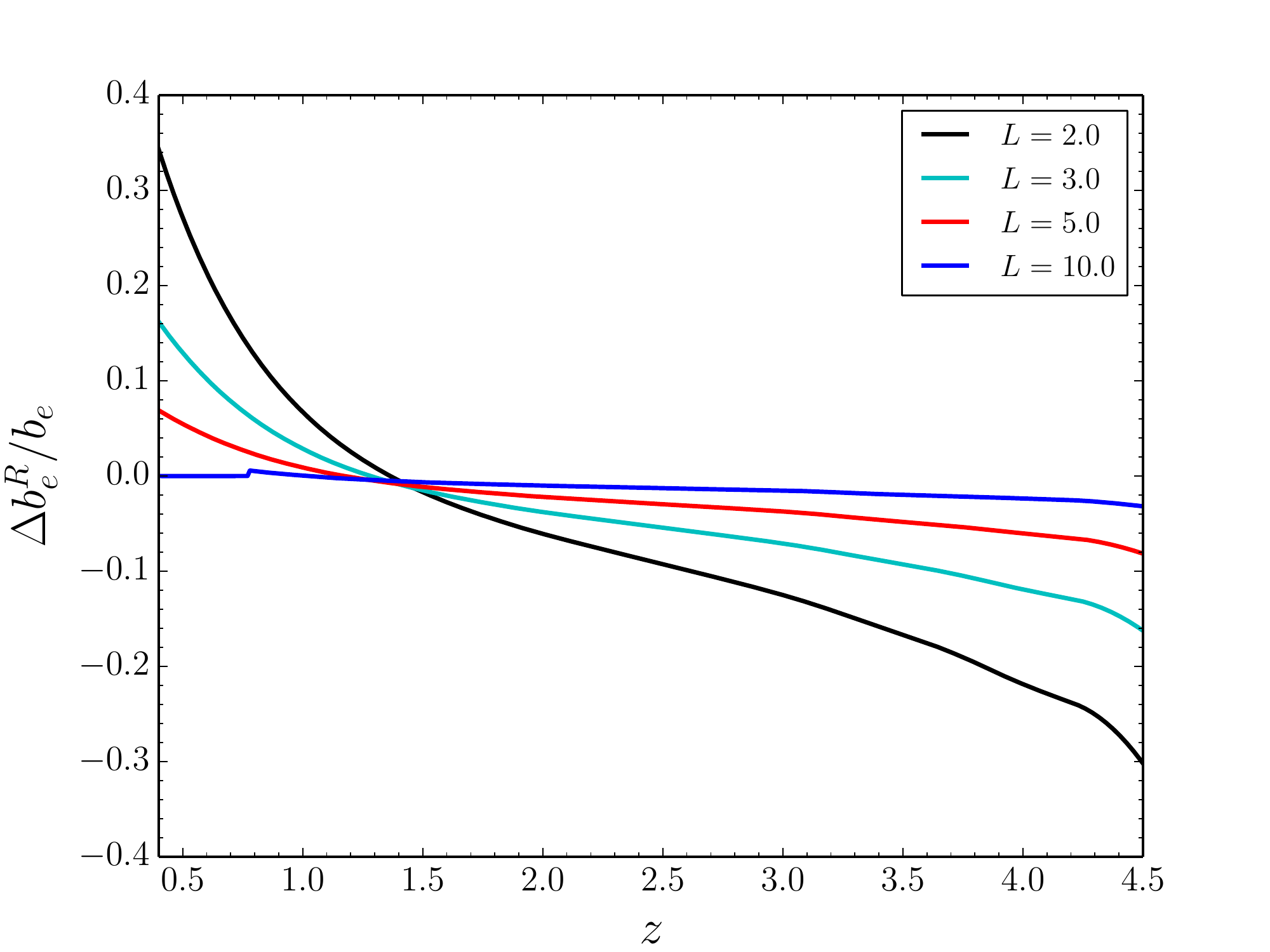}
\caption{Fractional difference  in the evolution bias due to nonlinear effects  for different values of the smoothing scale $L$ (in units of $h^{-1}\,$Mpc).  }
\label{fig:evobias}
\end{figure}

We show the fractional difference, $\Delta b^R_e /b_e= (b_e^R - b_e)/b_e$, in Fig. \ref{fig:evobias}, for various values of $L$. For $ L  =  10{h}^{-1}\,$Mpc, the difference is $\sim1\%$ and  for $L \gg 10{h}^{-1}\,$Mpc, we recover the FLRW limit as expected. The difference grows as $L$ is decreased below $10{h}^{-1}\,$Mpc. For $L\lesssim 2{h}^{-1}\,$Mpc, we need to go beyond second-order perturbation theory. The crossover point at $z\sim 1.5$ arises due to competititon between the nonlinear HI bias and the nonlinear lensing effect.
The nonlinear  bias  in  (\ref{renormevo}) is negative and the lensing contribution ${T}_{\kappa}$ is positive, and they cancel  at $z \sim 1.5$.
The fractional difference is dominated by the nonlinear bias for $ z \lesssim 1.5$ and for $z \gtrsim 1.5$ by the nonlinear lensing effect.

The kernels in \eqref{eq:z2} are as follows.
$G_2 ({\k}_1,{\k}_2)$  is the kernel for peculiar velocity at second order, which we approximate by its form in a matter-dominated universe:
\begin{eqnarray}\label{eq:G2}
 G_2({\k}_1,{\k}_2)= \frac{3}{7}+\frac{1}{2} \left({k_1 \over k_2}+{k_2 \over k_1}\right) {{\k}_1 \cdot {\k}_2 \over k_1
k_2} +\frac{4}{7}
\left({{\k}_1 \cdot {\k}_2 \over k_1
k_2}\right)^2.
\end{eqnarray}
The ${\cal K}$ kernels encode the nonlinear effects from mode coupling: ${\cal K}_{\rm R}$ -- RSD (velocity-velocity and velocity-density);  ${\cal K}_{\rm L}$ -- lensing (lensing-density and lensing-Kaiser); ${\cal K}_{\rm T}$ -- time perturbation (velocity-density and velocity- velocity). These are given by:
\begin{eqnarray}
\mathcal{K}_{\text{R}}({\k}_1,{\k}_2)&=&{b_1    f}\left[ \mu_1^2 + \mu_2^2 +\mu_1\mu_2\left({k_1 \over k_2}+{k_2 \over k_1}\right) \right] 
+ f^2 \left[ 2\mu_1^2 \mu_2^2 + {\mu_1\mu_2}\left( \mu_1^2 \frac{k_1}{k_2}+ \mu_2^2\frac{k_2}{k_1}\right) \right]\,,
\\
\mathcal{K}_{\text{L}}({\k}_1,{\k}_2)&=& 3\sqrt{1-\mu^2_1}\sqrt{1-\mu_2^2}\bigg[\left({k_1 \over k_2}+{k_2 \over k_1}\right)
b_1   + f\left(\frac{k_1\mu^2_1}{k_2}+\frac{k_2\mu^2_2}{k_1}\right)
 \bigg]T_{\kappa}\,,
 \\
\mathcal{K}_{\text{T}}({\k}_1,{\k}_2)&=&-  (1+z)f\HH
 \left[b_1 f+\frac{\d b_1}{\d z   }+\left(\mu^2_1+\mu_2^2\right)\left(f+ \frac{\d}{\d z}\ln (f\HH)\right)\right]\left(\frac{\mu_1}{k_1}+\frac{\mu_2}{k_2}\right),
\end{eqnarray}
where
\begin{eqnarray}
\mu_{a} = {{\k}_a \cdot {\n}\over k_a},\quad a=1,2,\dots
\end{eqnarray}

The  power spectrum of observed HI temperature fluctuations  in a distant observer approximation then follows from \eqref{eq:z1} and \eqref{eq:z2}:
\begin{eqnarray}\label{eq:powerSpec}
{P}_{T}({k},\mu) &=&\big( \bar{T}\big)^2 \left[b _1 +f\mu^2+\mathcal{A}\,\frac{\HH^2}{k^2}+i \mu\mathcal{B}\, \frac{  \HH}{k}\right]\left[b _1 +f\mu^2+\mathcal{A}\,\frac{\HH^2}{k^2}-i \mu\mathcal{B}\, \frac{  \HH}{k}\right]P_{m}(k) \nonumber\\
&& +\frac{\big( \bar{T}\big)^2}{2}\int \frac{\d^3k_1}{(2\pi)^3} \bigg[ b_2+b_1    F_2({\k}_1,{\k}-{\k}_1) + f \mu^2 G_2 ({\k}_1,{\k}-{\k}_1) \nonumber\\
&&
  + \mathcal{K}_{\text{R}}({\k}_1,{\k}-{\k}_1)
+\mathcal{K}_{\text{L}}({\k}_1,{\k}-{\k}_1)
+i\mathcal{K}_{\text{T}}({\k}_1,{\k}-{\k}_1)\bigg]^2
P_{m}(|{\k}-{\k}_1|)P_{m}({k}_1).
\end{eqnarray}
Once again, this power spectrum has a nonzero constant-power limit as $k\to 0$:
\begin{eqnarray}\label{eq:Tbnoisepower}
{P}_{T} \to
\frac{\bar{T}^2}{2}\int \frac{\d^3k_1}{(2\pi)^3}\Big[b_2     +3(1-\mu^2_1)\left( b _1 + \mu^2_1 f\right)T_{\kappa}\Big]^2 P^2_{m}({k}_1).
\end{eqnarray}

The monopole of the power spectrum is
\begin{eqnarray}\label{eq:monopolepower}
P^{\ell=0}_T(k)&=& (\bar{T})^2\bigg[b_1^2 +\frac{2}{3} b_1  f+ \frac{1}{5}f^2+ \frac{1}{3} \left[\mathcal{B}^2
+ 2\left( 3b _1 + f \right) \mathcal{A}\right]\frac{\HH^2}{k^2}
+\mathcal{A}^2\frac{\HH^4}{k^4}\bigg]P_m(k) 
 +\frac{1}{2} \int_{-1}^{1}\d \mu\, P^{(2)}_{T}(k,\mu) ,
\end{eqnarray}
where $P^{(2)}_{T}$ is the second order part of  (\ref{eq:powerSpec}).

\section{Results and Discussion}\label{sec:discus}

There are two key nonlinear effects on the ultra-large scale power spectrum that we have identified -- modification of the evolution bias and constant power that mimics white noise on horizon scales. These could potentially affect the standard analysis on horizon scales which assumes that  linear theory is valid. 

Here we compute the nonlinear modifications  to the power spectrum $P_T$ of the renormalized HI temperature fluctuations, as given in (\ref{eq:powerSpec}) and (\ref{eq:monopolepower}). For this computation,
we compute  the HI bias parameters using a standard prescription in Appendix B.  We set the limits of the convolution integral in  (\ref{eq:powerSpec}) following \cite{Carlson:2009it}: $(k_{1}^{\rm min}, k_{1}^{\rm max})= (10^{-4},10^4)h^{-1}\,$Mpc.

 \begin{figure}[htb!]
\includegraphics[width=0.49\textwidth]{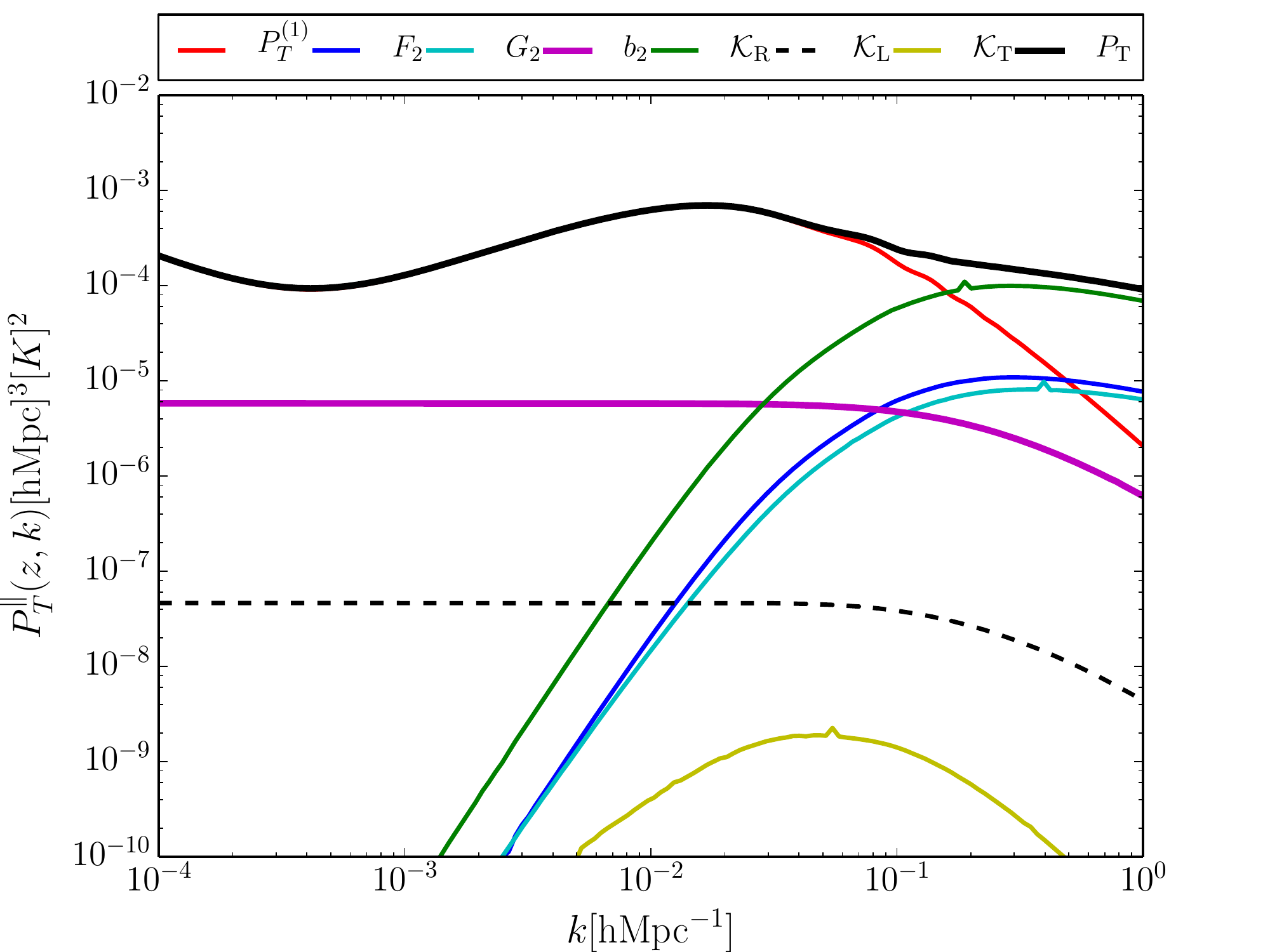}
\includegraphics[width=0.49\textwidth]{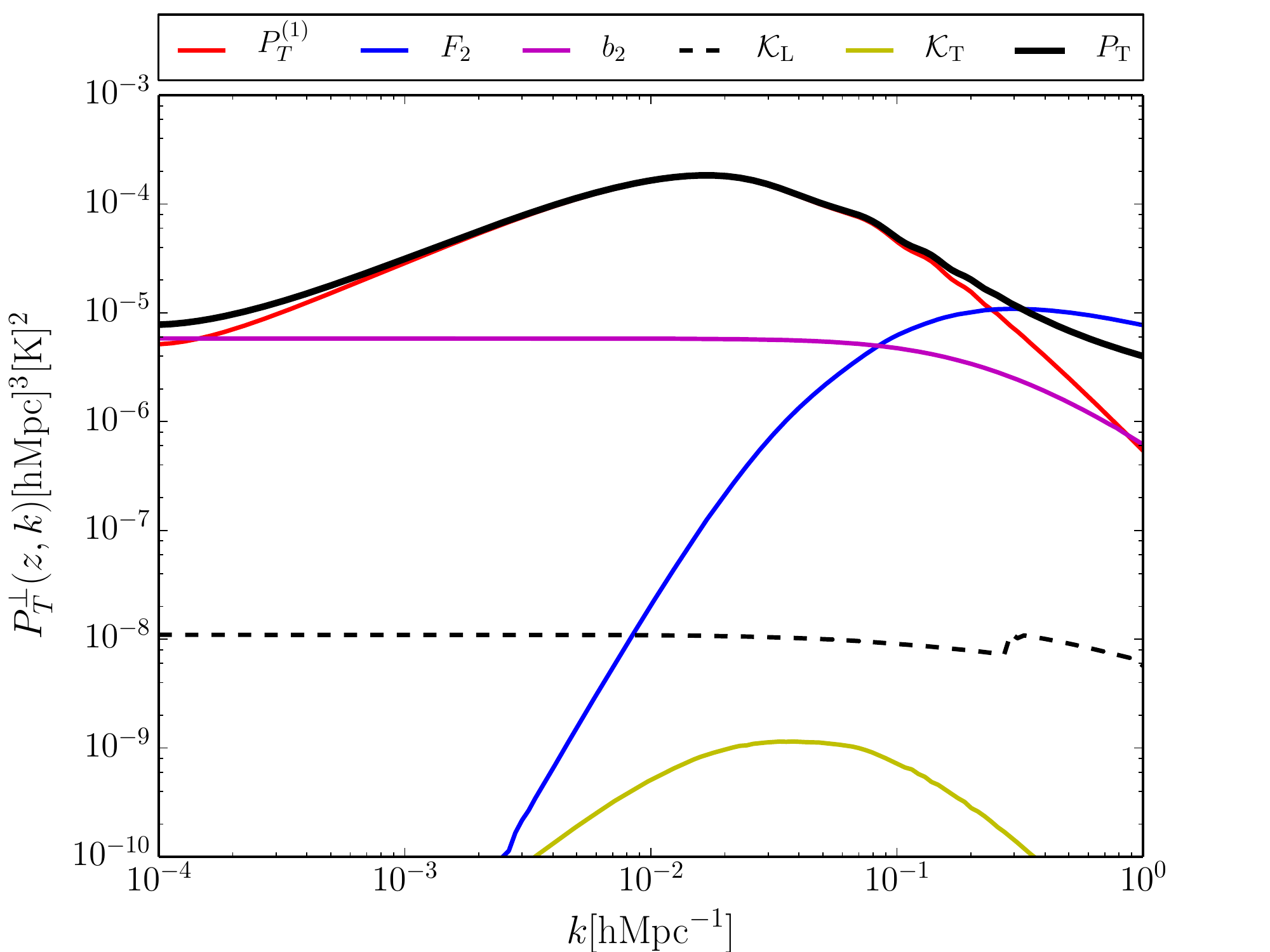}
\caption{Power spectrum $P_T$ of the renormalized  observed HI brightness temperature, at $z=0.5$. {\em Left:}  along the line of sight; {\em Right:} transverse to the line of sight. The solid black curve shows the total power and the red curve is the linear part $P_T^{(1)}$ of \eqref{eq:powerSpec}. The remaining curves show the nonlinear effects from: $F_2$ -- second-order density perturbations, $G_2$ --  RSD from second-order velocity,  $b_2$ -- HI bias,  $\mathcal{K}_{\rm R}$ -- RSD from mode coupling, $\mathcal{K}_{\rm L}$ -- lensing effect from mode coupling, $\mathcal{K}_{\rm T}$ -- time perturbation from mode coupling. 
}
\label{fig:powerspec1}
\end{figure}

In practice, the choice of the smoothing scale  $1/k_S$ will  be connected to the
resolution of the experiment. This scale cannot however be smaller than the scale $1/k_{\rm nl}$ if we wish to remain in the regime of validity of the perturbative approach. In our calculations we set $k_S\sim k_{\rm nl}\sim L^{-1}$.
 We note that although the results can be sensitive to the exact
value of $k_{\rm nl}$, this is not an issue as long as $1/k_S > 1/ k_{\rm nl}$. The
calculations will then only depend on $k_S$ but this is just a statement
that the power spectrum needs to be smoothed on the resolution scale
of the experiment in order to compare to observations. 

We do not use a window function to regulate  the convolution integral in (\ref{eq:powerSpec}). Careful analysis using a smoothed density field and angular power spectrum will be presented elsewhere.

\begin{figure}[htb!]
\includegraphics[width=0.490\textwidth]{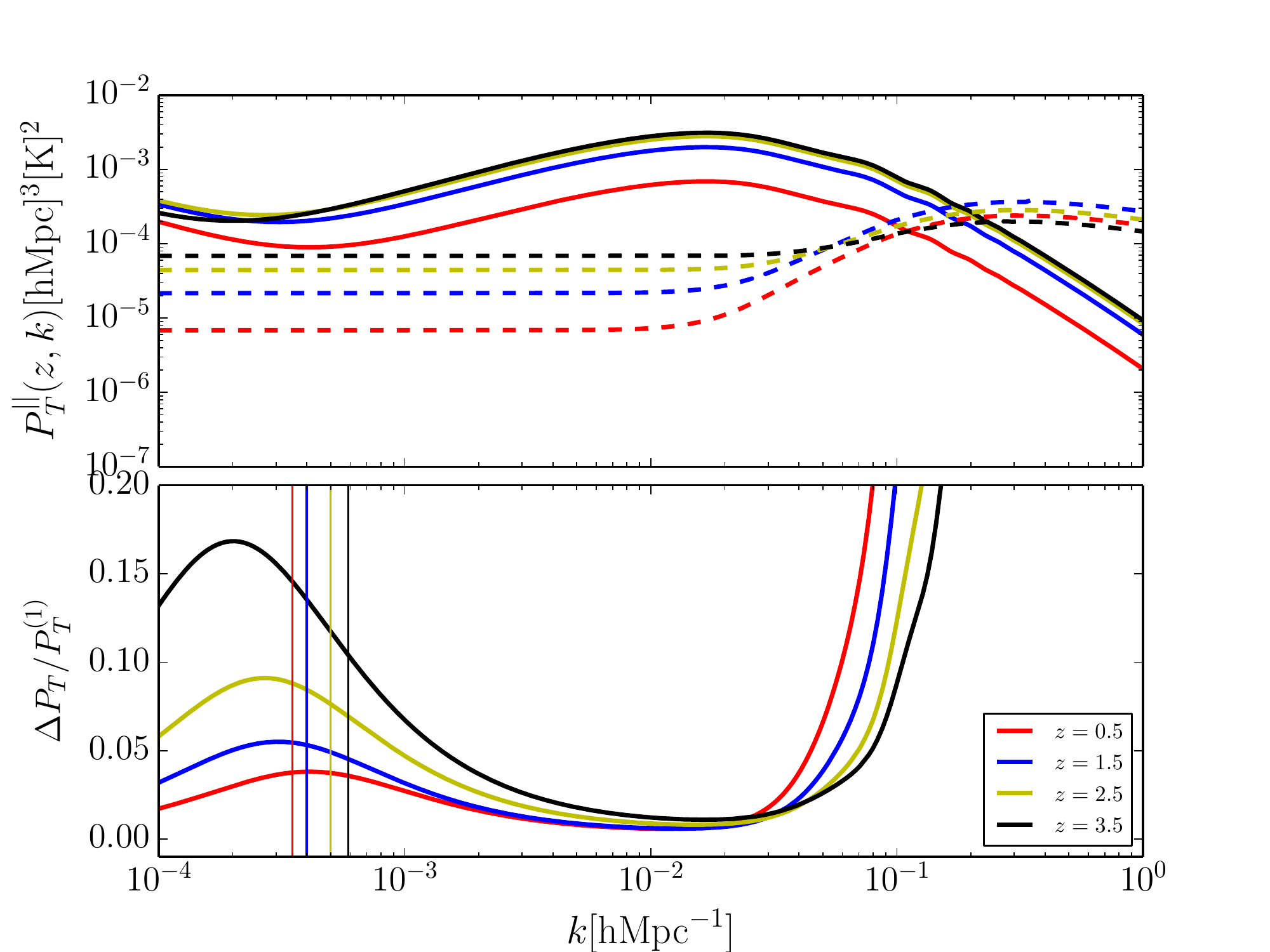}
\includegraphics[width=0.490\textwidth]{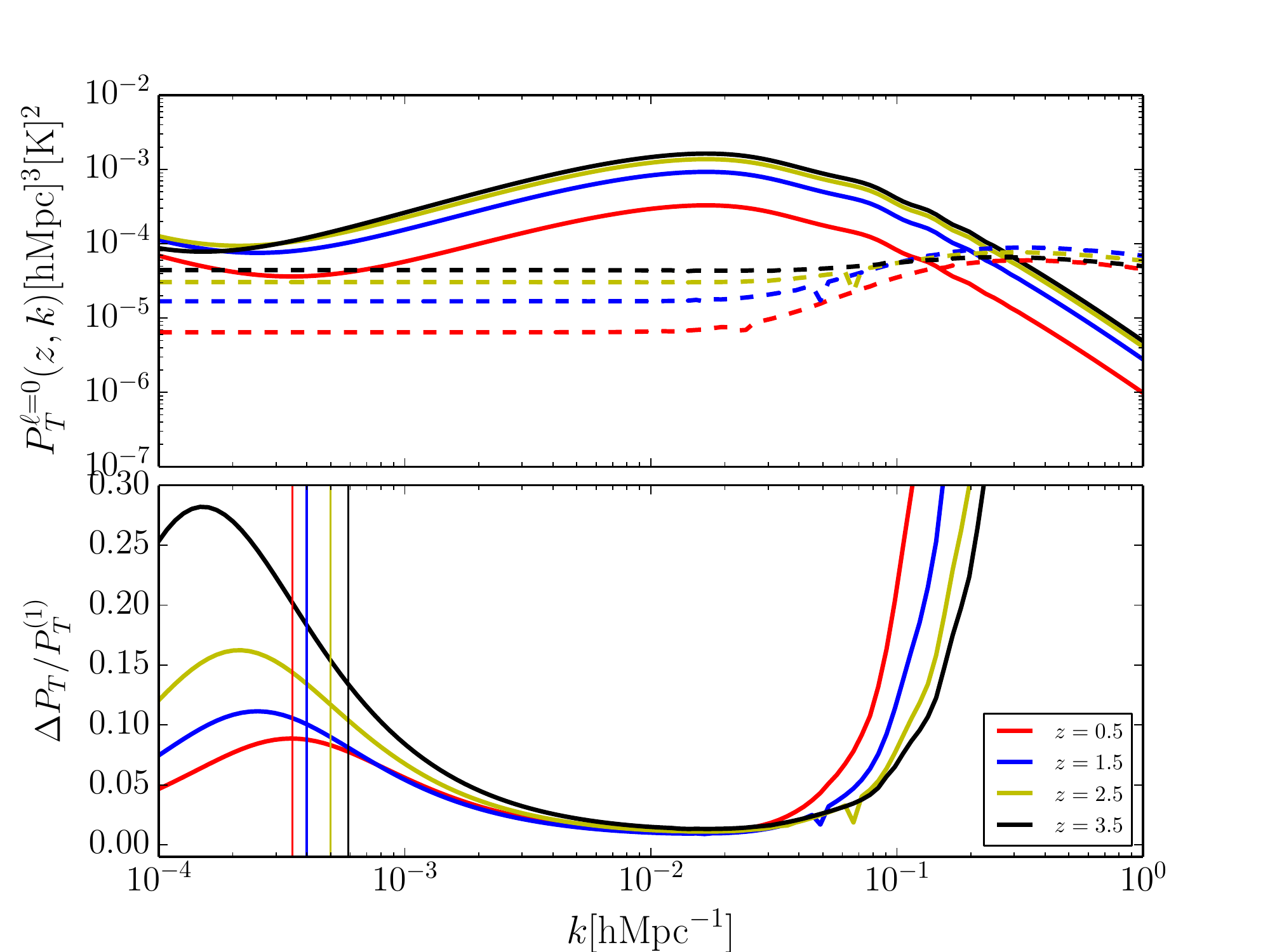}
\caption{Power spectrum $P_T$ of the renormalized observed HI brightness temperature at various redshifts. {\em Left panels:} along the line of sight; {\em right panels:} monopole.
{\em Upper panels:} Solid lines are the linear power spectrum, dashed lines the nonlinear power. {\em Lower panels:} The fractional difference due to nonlinear effects. Vertical lines indicate the comoving Hubble scale.  
}
\label{fig:Deltapk}
\end{figure}

Figure \ref{fig:powerspec1} illustrates  the amount of power contributed by the terms in \eqref{eq:powerSpec}, for $\mu=1,0$. The term responsible for non-vanishing of the average of $\Delta_T$ contributes constant power on large scales.
At linear order, the horizon-scale GR effects in $P_T^{(1)}$ become important near the horizon ($ k \sim \HH$) and dominate on super-Hubble scales ($k< \HH$). (We have not computed the horizon-scale GR effects at second order, since this in itself is a major task which is left for further work.)
Along the line of sight ($\mu = 1$), the
nonlinear RSD from mode coupling (via $\mathcal{K}_{\text{R}}$)  clearly dominates the nonlinear contribution $P_T^{(2)}$ at $k \gtrsim 0.05$. For  $ k \gtrsim 0.1$, it dominates over even the linear power spectrum $P_T\one$. At BAO scales, its contribution is $\sim10\%$.
As expected $\mathcal{K}_{\text{R}}$ vanishes for $\mu = 0$.



We quantify in Fig. \ref{fig:Deltapk} the total second-order effect on the linear power spectrum, focusing on large scales.
The fractional difference is defined as $\Delta P_{T}/P_T\one= \big[P_{T} -P_{T}\one\big]/P_{T}\one$. The contribution of the dominant constant power term peaks  around the  horizon scale $ k=\HH(z)$. The constant power from the lensing term $\mathcal{K}_{\text{L}}$  dominates over the constant power from nonlinear bias  at about $z \sim 1$ . 

     \begin{figure}[h!]
\includegraphics[width=120mm,height=70mm ]{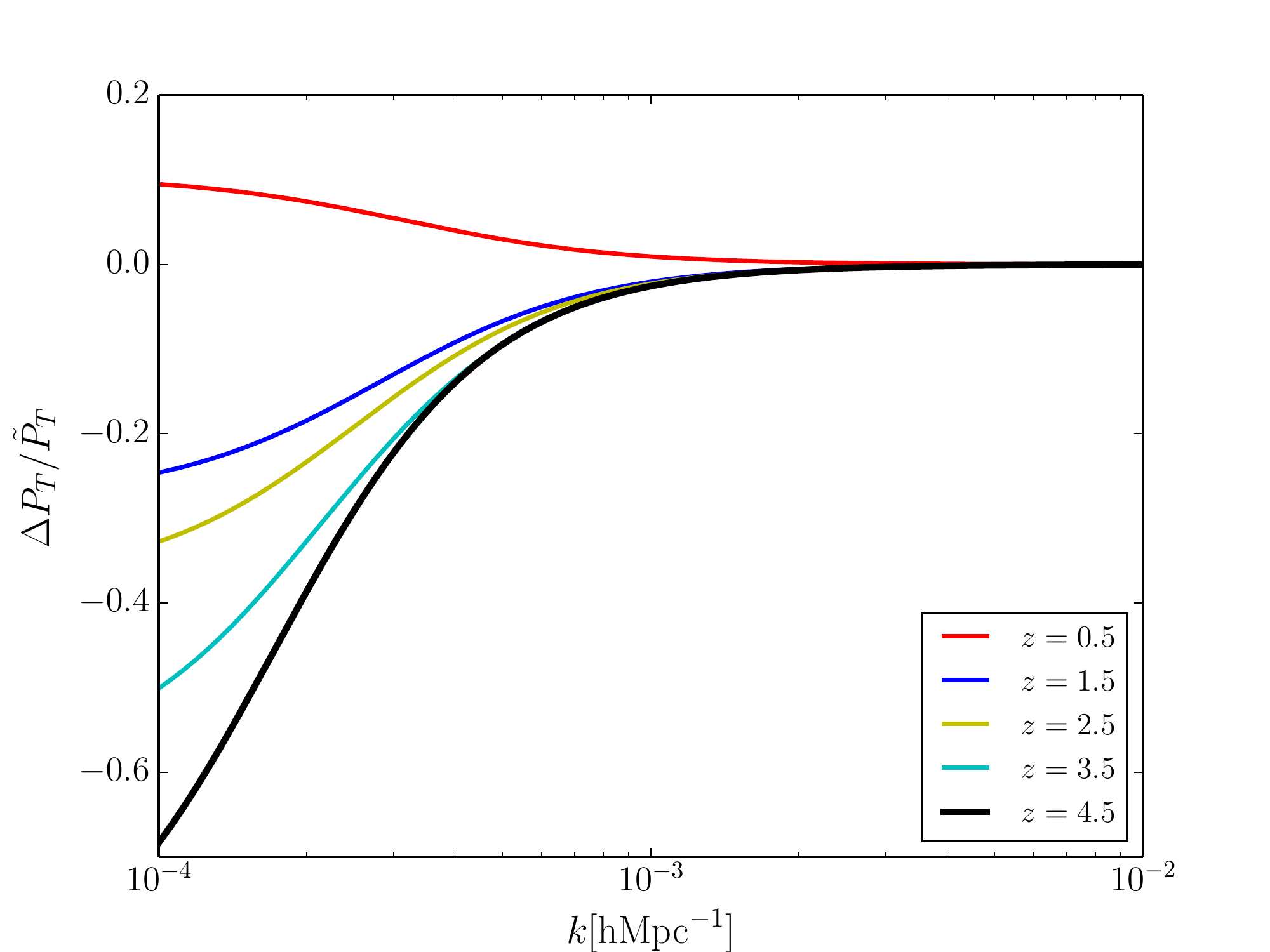}
\caption{Effect of renormalized evolution bias on the monopole of the power spectrum on large scales.  (The line of sight  power spectrum  is identical.) The smoothing scale for the evolution bias is set at $L= 5(1+z)^{-{2}/{(2+ n_s)}}h^{-1}\,$Mpc. }
\label{fig:EvoDeltapk}
\end{figure}

 We also  quantify the error on the power spectrum associated with using the bare evolution bias  (\ref{eq:evolbias}),  instead of the renormalized evolution bias  (\ref{renormevo}). To this end, we define the fractional difference as $\Delta P_{T}/\tilde{P}_{T} \equiv \big({{P_{T}} -\tilde{P}_{T}}\big)/{\tilde{P}_{T}}$, where $\tilde{P}_{T}$ is the power spectrum with bare evolution bias,  given by  (\ref{eq:evolbias}).  
Figure \ref{fig:EvoDeltapk}) shows that the difference could be more than 20\% on horizon scales.

Furthermore, what about third-order terms that we have neglected? These terms can contribute to the one-loop correction to the tree-level power spectrum through  cross-correlation with first-order terms. Here we argue that the effect of  third-order terms will be sub-dominant on the large scales that we are interested in.   The  brightness temperature up to third order is
      \begin{eqnarray}\label{eq:brightHI3}
T^{\rm obs}(z,{\n})&=&  \bar{T} (z)\Big[1+\Delta_{T}\one(z,{\n})+\frac{1}{2}\Delta\two_{T}(z,{\n})+ \frac{1}{3!}\Delta_T\three(z,{\n})\Big].
\end{eqnarray}
Assuming  Gaussian initial conditions, the main contribution to  the mean brightness temperature after renormalization will come from the third-order bias term. For a local Eulerian bias model, $\delta_{n} = b_1 \delta_m + b_2 \delta^2_m/2 + b_3 \delta^3_m/3!$, the ensemble average of the third-order term is given by
\cite{Bernardeau:2001qr}:
\begin{eqnarray}
b_3\big\langle \delta_m^3\big\rangle=3b_3 \big\langle \big(\delta_m\one\big)^2\delta_m\two\big\rangle= \frac{34}{7} b_3\big(\sigma^2_S\big)^2.
\end{eqnarray}
Using a Sheth-Torman mass function \cite{Sheth:1999mn,Sheth:2001dp}, we can show that $-1<b_3<1$ for $z \le 3.5$, as seen in Fig. \ref{fig:biasparameters} (see appendix \ref{sec:bias}).
 This implies that for $\sigma^2_{ S} < 1 $, we have $ (\sigma^2_{S})^2  < \sigma^2_{ S}$.
Therefore $\<\Delta\three_{T}\> < \<\Delta\two_{T}\>$, i.e the contribution from third-order fluctuations to the mean brightness temperature is sub-dominant.

 The mode coupling term  $\<\Delta_T\one(z,{\k})\Delta_T\three(z,{\k})\>$, will contribute to the  power spectrum at one-loop as follows:
 \begin{eqnarray}\label{eq:otheroneloop}
 P^{(13)}_T(k,\mu)&=&\frac{1}{3}\left(b _1 +f\mu^2+\mathcal{A}\,\frac{\HH^2}{k^2}+i \mu\mathcal{B}\, \frac{  \HH}{k}\right) P_m(k)\int \frac{\mathrm{d}^3k_1}{(2\pi)^3} P_m(k_1) \mathcal{K}\three( k,{\k}_1,-{\k}_1,\mu),
 \end{eqnarray}
 where $\mathcal{K}\three$ is the kernel for all third-order terms. 
 Although  (\ref{eq:otheroneloop}) is a one-loop contribution to the tree-level power spectrum, it does not appear as a convolution of two power spectra, i.e.
it does not produce a coupling of  power on small scales to ultra-large scales that can modify our results on those large scales. Also if we set $k_1 =k r$,  and take the limit $k \rightarrow 0$, the pre-factor in  (\ref{eq:otheroneloop}) goes to zero faster than the rate at which the $r$-integral goes to larger values.

Finally, it is important to point out  that the expression for  HI brightness temperature given in equation (\ref{eq:arbbrightness}) is a special limit of the HI brightness temperature during the Epoch of Reionization (EoR): 
\begin{eqnarray}\label{eq:EoR}
\delta T  = \bar{T} \bar{x}_{\text{\tiny{HI}}}\left( 1 +\delta_n\right)\left(1+ \frac{\delta{x_{\text{\tiny{HI}}}}}{\bar{x}_{\text{\tiny{HI}}}}\right) \left( 1 +\frac{\delta{J}}{\bar{J}}\right) \left( 1 - \frac{T_{\text{\tiny{CMB}}}}{T_s}\right)
\end{eqnarray}
where $\bar{x}_{\HI}$ is neutral fraction of hydrogen, $\delta x_{\HI}$ is the perturbation in neutral fraction of hydrogen, $T_{\text{\tiny{CMB}}}$ is the temperature of the CMB, $\delta J$ is the perturbed Jacobian and $T_s$ is the spin temperature. Equation (\ref{eq:EoR}) reduces to  equation (\ref{eq:arbbrightness}) in the limit $T_{\text{\tiny{CMB}}} \ll T_s$.  Despite these differences, the modulation of the HI power spectrum on ultra-large scales by nonlinear effects we discussed  at low-z will also be applicable during the EoR because the dominant nonlinear terms come from the perturbed Jacobian ($\delta J$) and not from $T_s$ and $T_{\text{\tiny{CMB}}}$ which we neglected.

\section{Conclusion}\label{sec:conc}

We derived for the first time the second-order perturbation of the HI brightness temperature in the post-reionization universe, 
focusing on the dominant  nonlinear redshift space distortion terms and the nonlinear lensing term.
 We find that:
\begin{itemize}
\item  The physical mean density of HI or the mean brightness temperature is modified by nonlinear terms that depend on redshift and the cutoff scale. By smoothing over structures on scales where perturbation theory may not be trusted, we showed that this could lead to more than 20\% change to the power spectrum on large scales.

\item The constant power on large scales associated with nonlinear evolution of  biased tracers can significantly affect the power spectrum on  horizon scales.
The boost in amplitude could affect the signal to noise ratio for possible detection of general relativistic effects and primordial non-Gaussianity from intensity mapping.

 \end{itemize}
 
However, the power spectrum is not observable. In order to determine whether these effects do lead to observable consequences, we need to compute the observed angular power spectrum $C_\ell(z,z')$. This is a subject for further investigation.

\acknowledgments  
We thank Daniele Bertacca, Ruth Durrer and Alex Hall for useful comments. OU thanks Chris Clarkson,  Didam Duniya and Bishop Mongwane for extensive discussions.
We  are supported by the South African National Research Foundation and South African Square Kilometre Array Project.   RM also acknowledges support from the UK Science \& Technology Facilities Council (grant ST/K0090X/1).


\appendix

\section{Brightness temperature in perturbation theory}\label{perturb}

Here we provide details for the perturbation calculation up to second order, using the approach developed in \cite{Umeh:2012pn,Umeh:2014ana}.
We consider a perturbed FLRW spacetime in the Poisson gauge assuming  a flat background metric
\begin{eqnarray}\label{eq:metric}
\d \hat s^2 &=&a^2\d s^2=a^2\left[-(1 + 2\Phi+\Phi\two )\d \eta^2 + 2\omega_{i} \d x^{i}\d \eta
 + \left((1-2 \Psi-\Psi\two)\delta_{i j} + h_{ij}\right)\d x^{i}\d x^{j}\right]\,.
\end{eqnarray}
Here the  proper time is related to conformal time  by   $\d t= a\d \eta$.  $\Phi$ is the first-order Newtonian gravitational potential, $\Psi$ is scalar curvature perturbation, and $\omega_i=\omega_i\V$ and $h_{ij}=h_{ij}\T$ are the second-order vector and tensor contributions, their divergences vanish. Since the null structure is unchanged under conformal transformation,  the metric $\hat{g}_{ab}$ associated with the line element $\d \hat s^2$, i.e  physical spacetime  maps the null geodesic  of the physical spacetime $\hat g_{ab}$ to a null geodesic on a perturbed Minkowski space time $ g_{ab}$ with the affine parameter associated  with each metric,  transforming as $\d \hat \lambda \rightarrow \d \lambda = a^{-2} \d {\lambda}$. The photon 4-vector transforms as $\hat k^b=a^{-2}k^b\Leftrightarrow\hat k_a=k_a$. For the matter 4-velocity, we have $\hat u^a=a^{-1}u^a\Leftrightarrow \hat u_a=au_a$ .  Hence, the photon energy transforms as $\hat E= -\hat u_b\hat k^b = -a \,u_b k^b=a E$.  
For any  tensor $S$ we  expand it up to second order as
\begin{equation}\label{eq:stdprescription}
 \hat{S}=\bar{S}+\delta\one S +\frac{1}{2}\delta\two S\,.
\end{equation}
We expand the 4-velocity, $u^a$,  of a matter field using  (\ref{eq:metric}) up to second order
\begin{eqnarray}
u^0&=&1 - \Phi +  \frac{3}{2} \Phi^2 -  \frac{1}{2}\Phi\two + \frac{1}{2}\bar{\D}_{i}v\bar{\D}^{i}v\\
u^i&=&\bar{\D}^{i}v + \frac{1}{2}
v^{i}{}\two + \frac{1}{2} \bar{\D}^{i}v\two
\end{eqnarray}

The perturbed photon tangent vector is calculated from the geodesic  $k^a\nabla_a k^b = 0$ and the redshift is given  by
    \begin{eqnarray}\label{eq:reshiftexp}
1+\hat{z}=1+\delta z +\frac{1}{2}\delta\two z&=& \frac{k^au_a|_s}{k^bu_b|_o}
\end{eqnarray}
 The background conformal time or the conformal metric affine parameter is mapped to the  conformal time in redshift space according to
 \begin{equation}\label{eq:conformaltimetrans}
 \lambda = \nu +\delta \lambda +\frac{1}{2}\delta\two \lambda
 \end{equation}
 Hence the physical scale factor is related to the background  scale factor through Taylor series expansion 
  \begin{eqnarray}\label{perta}
 a(\lambda) 
&=& a(\nu_s)\left[1+ \HH_s\delta \lambda+\frac{1}{2}\left(\HH_s\,{\delta}\two\lambda  +\left( \HH'_s+ \HH_s^2\right) ( \delta \lambda)^2\right)\right]\,,
 \end{eqnarray} 
  where we have performed the Taylor series expansion around the position of the physical position of the source. For consistency,  we could  define a homogeneous  redshift in terms of the physical scale factor as
 \begin{eqnarray}\label{eq:redefiendredshift}
 (1+z_s)&=& \frac{a(\nu_o)}{a(\nu_s)}\,.
 \end{eqnarray}
 We have also made the following  replacements in  (\ref{perta}) 
  \begin{equation}
 \frac{1}{a}\frac{\d a}{\d \lambda}\bigg|_{\nu_s}= \HH_s, \qquad 
 \frac{1}{a}\frac{\text{d}^2a}{\text{d} \lambda^2}\bigg|_{\nu_s}=\left(\mathcal{H}'_s+ \mathcal{H}^2_s\right)\,.\,, 
 \end{equation}
Inverting   (\ref{eq:reshiftexp}) and  using  (\ref{perta}) leads to 
  \begin{eqnarray}\label{phya}
 \frac{1}{(1+z_s)}&=& \frac{a(\nu_s)}{a(\nu_o)}\left[1+\left(\HH_s\delta \lambda - \delta z\right)+ \frac{1}{2}\left(\HH_s \delta\two \lambda- \delta\two z + 2(\delta z)^2 - 2\HH_s \delta z \delta \lambda 
 + \left(\HH'_s + \HH^2_s\right)(\delta \lambda)^2\right)\right]\,.
 \end{eqnarray}
Thus from  (\ref{phya}), the background scale factor may now be re-defined in terms of the physical redshift using  (\ref{eq:redefiendredshift}) and requiring that terms vanish at every order leads to 
 \begin{eqnarray}\label{eq;consistency}
 \delta \lambda &=&\frac{\delta z}{  \HH_s}\,,\\
 \delta\two\lambda &=& \frac{1}{\HH_s}\left[\delta\two z -(\delta z)^2 \left(1+ \frac{\HH'_s}{\HH^2_s}\right)\right]\,.
 \end{eqnarray}
 For further details on this see \cite{Umeh:2014ana}.
In order to expand the Jacobian in  (\ref{eq:brightness}), we need to relate 
the conformal time to the affine parameter associated with the physical metric according to ($\d \eta\rightarrow \d\lambda$)
     \begin{eqnarray}
     \frac{\d\lambda}{\d{\hat\lambda}}&=& \frac{1}{a^2(\eta)}\left[1+\delta k^0+\frac{1}{2}\left(\omega_{\p}+\delta\two k^0\right)\right]
     \end{eqnarray}
 and using  (\ref{eq:reshiftexp})  we link it to the physical redshift 
     \begin{eqnarray}\label{eq:lamtohatlam}
      \frac{\d\lambda}{\d{\hat\lambda}}&=&(1+\hat{z}_s)^2\left[1+\left(\delta k^0-2\delta z\right)+\frac{1}{2}\left(\omega_{\p}+\delta\two k^0-2\delta\two z+ 2\delta z\left( 3\delta z-2\delta k^0\right)\right)\right]
     \end{eqnarray}
 Differentiating  (\ref{eq:reshiftexp}) wrt $\lambda$ gives
     \begin{eqnarray}
     \frac{\d \hat{z}}{\d\lambda}&=& -\frac{\HH(\eta)}{a(\eta)}\left[1+\delta z+\frac{1}{2}\delta\two z -\frac{1}{\HH(\eta)}\left(\frac{\d\delta z}{\d\lambda}+\frac{1}{2}\frac{\d\delta\two z}{\d\lambda}\right)\right]
     \end{eqnarray}
 It is convenient  now to re-express background $\HH(\eta)$ and $a(\eta)$ in terms of the observed redshift 
     \begin{eqnarray}
        \frac{\d \hat{z}}{\d\lambda}&=& -\frac{\HH(\nu)}{a(\nu)}\bigg\{1
        +\left[\delta z-\frac{1}{\HH_s}\frac{\d  \delta z}{\d\lambda}-\left(\HH_s-\frac{\HH'_s}{\HH_s}\right)\delta \lambda\right]
        +
         \frac{1}{2}\left[\delta\two z -\frac{1}{\HH_s}\frac{\d\delta\two z}{\d\lambda}        
         -\left(\HH_s-\frac{\HH'_s}{\HH_s}\right)\delta\two\lambda
         \right.\\ \nonumber&&\left. 
        +2\delta \lambda \left(\frac{\d\delta z}{\d\lambda}-\delta z\left(\HH_s-\frac{\HH'_s}{\HH_s}\right)\right)
        +(\delta\lambda)^2\left(\HH_s^2-3\HH'_s+\frac{\HH''_s}{\HH_s}\right)\right]
       \bigg\}
     \end{eqnarray}
     Using  (\ref{eq:lamtohatlam}), we  substitute for $\d\lambda$ and after some simplification we find
      \begin{eqnarray}\label{eq:dzbydhatlam}
        \frac{\d \hat{z}}{\d{\hat\lambda}}&=& -\frac{\HH(\nu)}{a(\nu)^3}\bigg\{1+\left(\delta k^0 -\delta z -\frac{1}{\HH_s}\frac{\d\delta z}{\d\lambda}-\left(\HH_s-\frac{\HH'_s}{\HH_s}\right)\delta \lambda\right)
        +\frac{1}{2}\left[\delta\two k^0+\omega_{\p}-\delta\two z-\frac{1}{\HH_s}\frac{\d\delta\two z}{\d\lambda}
\right.\\ \nonumber&&\left.        
        -\left(\HH_s-\frac{\HH'_s}{\HH_s}\right)\delta\two\lambda+2\left((\delta z)^2-\frac{2}{\HH_s}\delta k^0\frac{\d\delta z}{\d\lambda}\right)+2\delta z\left(\frac{2}{\HH_s}\frac{\d\delta z}{\d\lambda} -\delta k^0\right)
        +2\bigg(\frac{\d\delta z}{\d\lambda}+\delta z\left(\HH_s-\frac{\HH'_s}{\HH_s}\right)
        \right.\\ \nonumber&&\left.
        -2\delta k^0\left(\HH_s-\frac{\HH'_s}{\HH_s}\right)\bigg)\delta \lambda
        +(\delta \lambda)^2\left(\HH_s^2-3\HH'_s+\frac{\HH''_s}{\HH_s}\right)\right]
        \bigg\}
        \end{eqnarray} 
  The Jacobian in  (\ref{eq:brightness}) is then obtained by simply inverting  (\ref{eq:dzbydhatlam})
\begin{eqnarray}\label{eq:dhatlamdbyz}
     J\equiv  \bigg| \frac{\d{\hat\lambda}}{\d \hat{z}}\bigg|&=& \frac{a(\nu)^3}{\HH(\nu)}\bigg\{1+\left(\delta z-\delta k^0  +\frac{1}{\HH_s}\frac{\d\delta z}{\d\lambda}+\left(\HH_s-\frac{\HH'_s}{\HH_s}\right)\delta \lambda\right)
       +\frac{1}{2}\left[\delta\two z-\delta\two k^0-\omega_{\p}+\frac{1}{\HH_s}\frac{\d\delta\two z}{\d\lambda}        
         \right.\\ \nonumber&&\left.      
        +\left(\HH_s-\frac{\HH'_s}{\HH_s}\right)\delta\two\lambda
        +\delta \lambda\left[2\frac{\d\delta z}{\d\lambda}\left(1-2\frac{\HH'_s}{\HH^2_s}\right)+2\delta z\left(\HH_s-\frac{\HH'_s}{\HH_s}\right)-2\delta k^0\left(\HH_s-\frac{\HH'_s}{\HH_s}\right)\right]
        +
         \right.\\ \nonumber&&\left.
        (\delta \lambda)^2\left(\HH^2_s-\HH'_s+2\left(\frac{\HH'_s}{\HH_s}\right)-\frac{\HH''_s}{\HH_s}\right) 
        -2\delta k^0\left(\delta z+\frac{1}{\HH_s}\frac{\d\delta z}{\d\lambda}\right)+2 \left((\delta k^0)^2 + \left(\frac{1}{\HH_s}\frac{\d\delta z}{\d\lambda}\right)^2\right)
        \right]
        \bigg\}
        \end{eqnarray}
 For the  HI number density we follow the standard prescription given in      (\ref{eq:stdprescription})to perturb it up to second order
     \begin{equation}\label{eq:perturbednumber}
n (\eta,x^i)
=\bar{n} (\eta)\left[1+ \delta_n  (\eta,x^i) +\frac{1}{2}\delta\two_n (\eta,x^i) \right]\,,
 \end{equation}
The background number density is then mapped to its equivalent in redshift  space
\begin{eqnarray}\label{eq:pertn}
 \bar{n} ({\eta}) 
&=& \bar{n}(\nu_s)\left[1+ \frac{\bar{n}'}{\bar{n}}\delta \lambda+\frac{1}{2}\left(\frac{\bar{n}'}{\bar{n}}\,{\delta}\two\lambda  +\frac{\bar{n}''}{\bar{n} } ( \delta \lambda)^2\right)\right]\,.
 \end{eqnarray}
The rate of change of background number density along the line of sight  define parameters that quantify our insufficient knowledge of the background spacetime, i.e the evolution bias parameters:
 \begin{eqnarray}
 b_e& =& \frac{\d \ln (a^3 \bar{n} (\bar{\eta}))}{\d\eta}=3\frac{a'}{a}+\frac{\bar{n}'}{\bar{n}}\,,\\
 \frac{\d b_e}{\d\lambda} &=& -\left(b_e-3\HH\right)\left(3\HH-b_e\right)
 +3\HH'+\frac{\bar{n}''}{\bar{n} }\,.
  \end{eqnarray}
  Putting  (\ref{eq:dzbydhatlam}) and  (\ref{eq:perturbednumber})  into  (\ref{eq:brightness}), we arrive at equations \eqref{eq:firstdeltaTb1in} and \eqref{eq:seconddeltaTb1in}.

Using the photon geodesic  $k^a \nabla_a k^b =0$, we calculate the weak lensing terms from the perturbed photon 4-vector, which at first order is given by
\begin{eqnarray}\label{eq:firstorderk}
\delta k^0(\lambda_s)&=& - 2(\Phi_s-\Phi_o)
   +\int_{\lambda_s}^{\lambda_o} (\Phi'+\Psi')\d \lambda\\
      \label{eq:firstorderk2} 
 \delta k_{\|}(\lambda_s)&=& -(\Phi_s-\Phi_o)+(\Psi_s-\Psi_o)
   - \int_{\lambda_s}^{\lambda_o}(\Phi' + \Psi' )\d \lambda\\
 \label{eq:firstorderk3}
 \delta k^i_{\bot}(\lambda_s)&=& -\int_{\lambda_s}^{\lambda_o} \nabla^i_{\bot}(\Phi+  \Psi)\d \lambda\,,
\end{eqnarray}
where we have set the perturbation of the geodesic at the observer to zero, $\delta k^a(\lambda_o)=0$. The radial component is  obtained by projecting on $e^i$, while the transverse component comes from projecting with the screen space metric, $N_{ij}= \bar{g}_{ij}+u_iu_j-e_ie_j$.
The perturbed position of the photon at first order is obtained by solving the following equation
 \begin{equation}\label{eq:pertposition}
\frac{\d\delta x^b}{\d\lambda} = \delta k^b -\delta k^c\partial_c x^b\,.
\end{equation}
Using equations (\ref{eq:firstorderk})--(\ref{eq:firstorderk3}) in  (\ref{eq:pertposition}) we find 
  \begin{eqnarray}
   \delta \lambda &=&\frac{\delta z}{\HH_s}=\frac{1}{\HH_s}\left[(\partial_{\p}v_{s}-\partial_{\p}v_{o})- (\Phi_s- \Phi_o) 
+\int_{\lambda_o}^{\lambda_s} (\Phi' + \Psi')\d \lambda\right]
\,,\\ 
   \Delta x_{\|}&=&-\left(\Psi_o+\Phi_o\right)\left(\lambda_o-\lambda_s\right)
   +\int^{\lambda_s}_{\lambda_o}
  \left(\Phi+\Psi\right)\d\lambda
  -\frac{1}{\HH_s}\left[(\partial_{\p}v_{s}-\partial_{\p}v_{ o})- (\Phi_s- \Phi_o) 
+\int_{\lambda_o}^{\lambda_s} (\Phi' + \Psi')\d \lambda\right], \\ \nonumber&&
 \\
  \Delta x_{\bot}^i&=&-\int_{\lambda_o}^{\lambda_s}(\lambda_s-\lambda) (\nabla^i_{\bot}\Phi+ \nabla_{\bot}^i \Psi)\d \lambda
  =\int_{0}^{\chi_s}(\chi-\chi_s) (\nabla^i_{\bot}\Phi+ \nabla_{\bot}^i \Psi)\d \chi\,.
   \end{eqnarray}

   Then the first order brightness temperature fluctuation \eqref{eq:firstdeltaTb1in} is given by   
     \begin{eqnarray}\label{eq:HIoverdensity}
      \Delta_{T}\one(z,{\n})&=&\delta_n  -\frac{1}{\HH_s}\partial^2_{\p}v_s+
      \frac{1}{\HH_s}\left(b_e-2\HH_s-\frac{\HH'_s}{\HH_s}\right)\partial_{\p}v_{s}+\frac{1}{\HH_s}\Psi'_s-\frac{1}{\HH_s}\left(b_e-3\HH_s-\frac{\HH'_s}{\HH_s}\right)
     \Phi_s
     \\ \nonumber&&
     +\frac{1}{\HH_s}\left(b_e-2\HH_s-\frac{\HH'_s}{\HH_s}\right)
\int_{\lambda_o}^{\lambda_s} (\Phi' + \Psi')\d \lambda +\Delta_{T}\one(z,{\x}_o)
     \end{eqnarray}
 where  $\Delta_{T}\one(z,{\x}_o)$ is a collection of all the terms measured at observer
     \begin{eqnarray}\label{eq:observerterms}
      \Delta_{T}\one(z,{\x}_o)&=&-\frac{1}{\HH_s}\left(b_e-\HH_s-\frac{\HH'_s}{\HH_s}\right)\partial_{\p}v_{ o}+\frac{1}{\HH_s}\left(b_e-3\HH_s-\frac{\HH'_s}{\HH_s}\right)\Phi_{ o}
     \end{eqnarray}

     Recently, the second-order large-scale perturbations to the observed galaxy number counts have been computed by
\cite{Bertacca:2014dra,Bertacca:2014wga,Yoo:2014sfa,DiDio:2014lka,Bertacca:2014hwa}. These results do not cover the case of HI intensity mapping. Substituting for all the terms in the second-order fluctuation (\ref{eq:seconddeltaTb1in}) leads to an unmanageably long expression. 
      We take a consistent extension of the Kaiser approximation \cite{Kaiser:1987qv} from first to second order, so as to reduce the complexity of the calculation. The full calculation will be presented elsewhere.
Obtaining a consistent Kaiser limit up to second order requires that we set 
      \begin{eqnarray}\label{eq:kaizerlimit}
     \delta k^0 \longrightarrow 0\,,\quad
     \delta	k\p \longrightarrow 0\,,\quad \delta\two k^0 \longrightarrow 0\,,\quad
     \delta\two	k\p \longrightarrow 0\,,\quad
     \delta z \longrightarrow 0\,,\quad \delta\two z \longrightarrow 0\,.
     \end{eqnarray}
     These terms contain a Doppler term and gravitational potential terms, and products of these two terms.
  We do not set these terms to zero in the post-Born correction terms because post-Born correction terms already contain higher number of derivatives than other terms in the equation. Implementing  the approximation of  (\ref{eq:kaizerlimit}) in  equations \eqref{eq:firstdeltaTb1in} and \eqref{eq:seconddeltaTb1in}, we find  
     \begin{eqnarray}
     \Delta_{T}\one(z,\n)&\approx&\delta_n   +\frac{1}{\HH_s} \frac{\d\delta z}{\d\lambda}\,,\\
    \Delta_T\two(z,\n)&\approx&\delta_n\two
     + \frac{1}{\mathcal{H}} \frac{\d\delta\two z}{\d \lambda}   + 2 \left( \frac{1}{\mathcal{H}}\frac{\d\delta z}{\d\lambda}\right)^2     
     + 2\delta_n \left( \frac{1}{\mathcal{H}}\frac{\d\delta z}{\d\lambda}\right)
      +2\bigg[\frac{\delta z}{\mathcal{H}_s} {\Delta_{T_{b}}\one}' + \Delta x_{\p} \partial_{\p} \Delta_{T_{b}}\one + \Delta x_{\bot}^i\nabla_{\bot i} \Delta_{T_{b}}\one\bigg]\,.
\label{eq:Kaiser2}
     \end{eqnarray}     
The last set of terms in  (\ref{eq:Kaiser2}) clearly show how the effect of weak lensing
 appears in an HI intensity map. Its appearance is solely due to the
post-Born correction that must be included if one goes beyond linear order \cite{Hagstotz:2014qea}. 
 The derivative of the redshift with respect to $\lambda$  contains the redshift space distortion terms and they are given by 
     \begin{eqnarray}
\frac{\d\delta z}{\d\lambda}=\D{\p}v'-\D\p^2v+\D{\p}\Phi+\Phi'\approx -\partial^2_{\p}v\,,\qquad \frac{\d\delta\two z}{\d\lambda}\approx -\partial^2_{\p}v\two,
\end{eqnarray}
where we have implemented the derivatives of the conditions listed in  (\ref{eq:kaizerlimit}).

Considering only the dominant terms in each case and putting these terms together, we obtain finally the key result of  \eqref{eq:SimpKaiser1}. 

   \section{HI bias from halo bias}\label{sec:bias}

We calculate the bias parameters from a simple Sheth-Torman mass function \cite{Sheth:1999mn,Sheth:2001dp}:   
\begin{eqnarray}\label{eq:Multibias1}
b_1   &=&1+\average{\frac{(q\nu-1)}{\delta_c}+\frac{2 p}{\delta_c\left(1+(q\nu)^p\right)}}\,,\\
b_2   &=&\frac{8}{21}\left(b_1   -1\right)+\average{\frac{4\left(p^2+ \nu p q\right)-(q\nu-1)\left(1+(q\nu)^p\right)-2p}
{\delta^2_c \left(1+(q\nu)^p\right)}
 +\frac{1}{\delta_c^2}\left((q\nu)^2 - 2q\nu -1\right)}\,
\label{eq:Multibias2}\\
b_{3} &=& -\frac{236}{189}\left( b_{1}-1\right) -\frac{13}{7} \left(b_{2}- \frac{8}{21}\left( b_{1}-1\right)\right)
+  \bigg<-\frac{\left(3 + 3 \nu q + 3 \nu^2 q^2 - \nu^3 q^3\right)}{\delta_c^3}
\\ \nonumber &&
+ \frac{\left(8 p^3 + 12 p^2 \left( 1 +\nu q\right) + p \left( 6 \nu^2 q^2 -2 \right) \right) }{\delta_c^3 \left( 1 + 1+ (\nu q)^p\right)} 
+6\frac{\left(1+ 2\nu q -\nu^2 q^2\right)}{\delta_c^3} - 24 \frac{\left( p^2 + \nu pq\right)}{\delta_c^3\left( 1+ (\nu q)^p\right)}
\\ \nonumber &&
 -4\frac{(1-\nu q)}{ \delta_c^3} + 8 \frac{p}{\delta_c^3\left( 1+ (q\nu)^p\right)}\bigg>_{\rm M}\,,
\label{eq:Multibias3}
\end{eqnarray}
where $q=0.707$ and $p=0.3$ are obtained from a fit to numerical simulations. The peak height is related to the variance $\sigma(M)$ of the density field smoothed on mass scale $M$ by
$
\nu=\left({\delta_\text{c}}/{\sigma_\text{G}}\right)^2.
$
 The averaging notation is defined by 
\begin{equation}\label{eq:averagedef}
X  (z,{\x})=\average{X_h(z,{\x})}= \frac{\int_{M_{-}}^{M_{+}} \d M\left[ X_{h}(z,{\x},M)M_{\HI}  (M) n_h(z,{\x},M) \right] }{\int_{M_{-}}^{M_+} \d M\left[ M_{\HI}(M)
n_h(z,{\x},M)\right],
}
\end{equation}
where 
$M_{-}$ and $M_{+}$ are the lower and upper limits of masses, which are related to the limits of circular velocity of galaxies that could house HI.  These are obtained from the circular velocity constraint
\begin{equation}\label{eq:velvsmass}
v_{\text{circ}} = 30 \sqrt{1+z} \left(\frac{M}{10^{10}M_{\odot}}\right)^{1/3} ~{\rm km\,s}^{-1}\,,
\end{equation}
where we  assumed that only halos with circular velocities between $30 - 200\,$kms$^{-1}$ are able to host HI. This range of circular velocity is motivated by observation \cite{Bull:2014rha}.

\begin{figure}[htb!]
\includegraphics[width=120mm,height=70mm ]{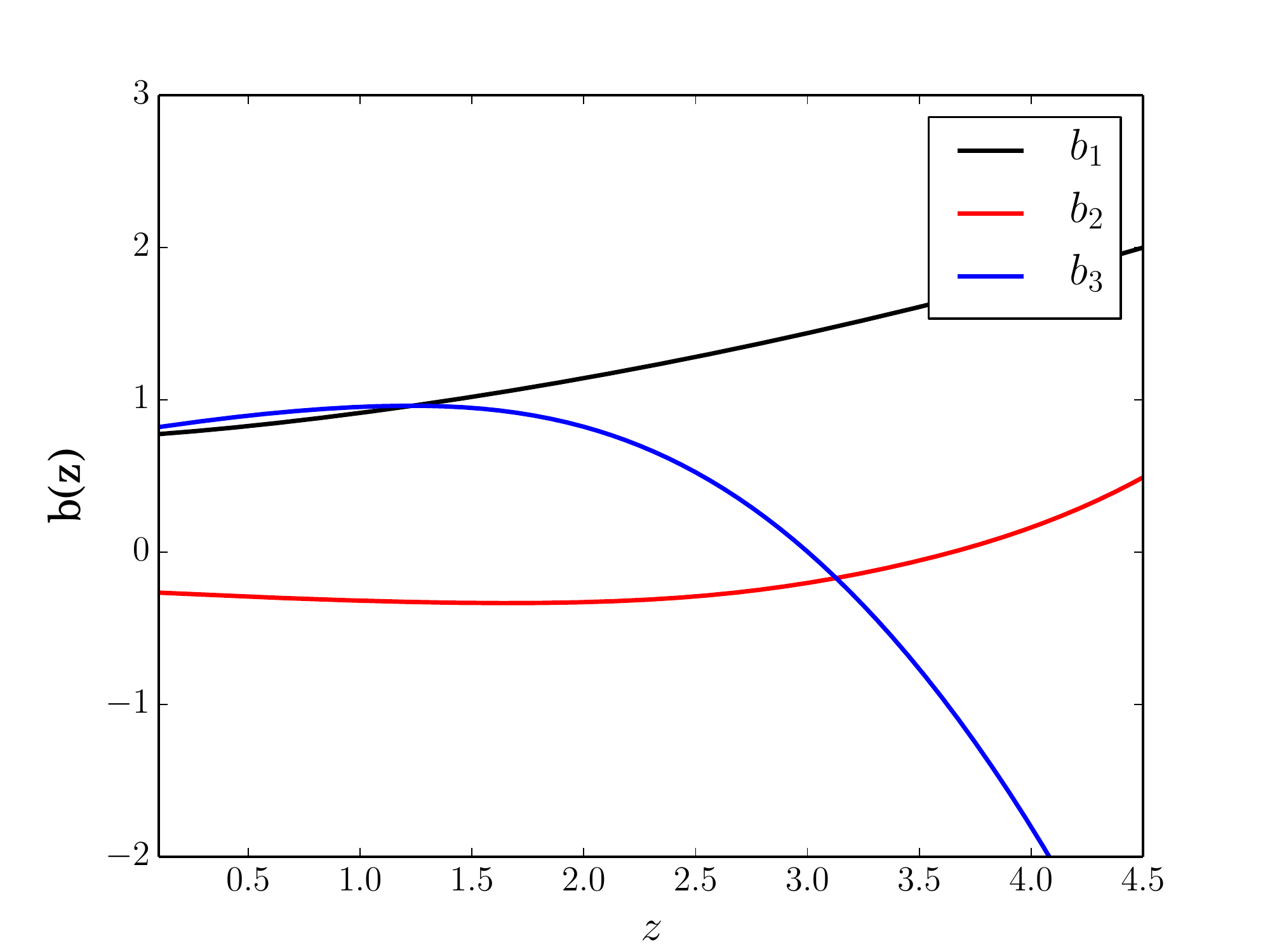}
\caption{The HI bias parameters used in the paper. }
\label{fig:biasparameters}
\end{figure}

We adopt a fitting function based on simulations for the  HI mass function \cite{Santos:2015gra}
\begin{equation}
M_{\HI}(M) = C M^{0.6}\,,
\end{equation}
where the normalization factor is chosen to match the measurement of $\Omega_{\HI}$ at $z =0.8$  \cite{Switzer:2013ewa}.
In general the local number of halos with mass $M$ is given by
\begin{equation}
n_h(M) = \nu f(\nu) \frac{\bar{\rho}}{M^2}\frac{\d\ln \nu}{\d\ln M},
\end{equation}
where the peak height $\nu$ is related to the non-gaussian dark matter variance, $\sigma_{nG}$, $\nu = (\delta_c/\sigma_{nG})^2$ and $\delta_c = 1.686$ is the threshold of linear density contrast. 
We show in Fig. \ref{fig:biasparameters} the shape of  each of the bias parameters given in equations (\ref{eq:Multibias1}-\ref{eq:Multibias3}).

\begin{thebibliography}{99}

\bibitem{Bull:2014rha} 
  P.~Bull, P.~G.~Ferreira, P.~Patel and M.~G.~Santos,
  Astrophys.\ J.\  {\bf 803},  21 (2015)
  [arXiv:1405.1452 [astro-ph.CO]].


\bibitem{Hall:2012wd} 
  A.~Hall, C.~Bonvin and A.~Challinor,
  Phys.\ Rev.\ D {\bf 87},  064026 (2013)
  [arXiv:1212.0728 [astro-ph.CO]].

\bibitem{Camera:2013kpa} 
  S.~Camera, M.~G.~Santos, P.~G.~Ferreira and L.~Ferramacho,
  Phys.\ Rev.\ Lett.\  {\bf 111}, 171302 (2013)
  [arXiv:1305.6928 [astro-ph.CO]].

\bibitem{Santos:2015gra} 
  M.~G.~Santos {\it et al.},
  arXiv:1501.03989 [astro-ph.CO].


\bibitem{Alonso:2015uua} 
  D.~Alonso, P.~Bull, P.~G.~Ferreira, R.~Maartens and M.~G.~Santos,
  arXiv:1505.07596 [astro-ph.CO].

\bibitem{Raccanelli:2015vla} 
  A.~Raccanelli, F.~Montanari, D.~Bertacca, O.~Dor\'e and R.~Durrer,
  arXiv:1505.06179 [astro-ph.CO]. 

\bibitem{Alonso:2015sfa} 
  D.~Alonso and P.~G.~Ferreira,
  arXiv:1507.03550 [astro-ph.CO].

\bibitem{Fonseca:2015laa} 
  J.~Fonseca, S.~Camera, M.~Santos and R.~Maartens,
  arXiv:1507.04605 [astro-ph.CO].


  
\bibitem{Baker:2015bva} 
  T.~Baker and P.~Bull,
  arXiv:1506.00641 [astro-ph.CO].

\bibitem{Smith:2002dz} 
  R.~E.~Smith {\it et al.} [VIRGO Consortium Collaboration],
  Mon.\ Not.\ Roy.\ Astron.\ Soc.\  {\bf 341}, 1311 (2003)
  [astro-ph/0207664].

  
\bibitem{Heavens:1998es} 
  A.~F.~Heavens, S.~Matarrese and L.~Verde,
  Mon.\ Not.\ Roy.\ Astron.\ Soc.\  {\bf 301}, 797 (1998)
  [astro-ph/9808016].

\bibitem{McDonald:2006mx} 
  P.~McDonald,
  Phys.\ Rev.\ D {\bf 74}, 103512 (2006)
  [Phys.\ Rev.\ D {\bf 74}, 129901 (2006)]
  [astro-ph/0609413].

\bibitem{McDonald:2009dh} 
  P.~McDonald and A.~Roy,
  JCAP {\bf 0908}, 020 (2009)
  [arXiv:0902.0991 [astro-ph.CO]].


\bibitem{Chan:2012jx} 
  K.~C.~Chan and R.~Scoccimarro,
  Phys.\ Rev.\ D {\bf 86}, 103519 (2012)
  [arXiv:1204.5770 [astro-ph.CO]].

\bibitem{Nishizawa:2012db} 
  A.~J.~Nishizawa, M.~Takada and T.~Nishimichi,
  Mon.\ Not.\ Roy.\ Astron.\ Soc.\  {\bf 433}, 209 (2013)
  [arXiv:1212.4025 [astro-ph.CO]].

\bibitem{Assassi:2014fva} 
  V.~Assassi, D.~Baumann, D.~Green and M.~Zaldarriaga,
  JCAP {\bf 1408}, 056 (2014)
  [arXiv:1402.5916 [astro-ph.CO]].

\bibitem{Ade:2015xua} 
  P.~A.~R.~Ade {\it et al.} [Planck Collaboration],
  arXiv:1502.01589 [astro-ph.CO].

\bibitem{Bertacca:2014dra} 
  D.~Bertacca, R.~Maartens and C.~Clarkson,
  JCAP {\bf 1409},  037 (2014)
  [arXiv:1405.4403 [astro-ph.CO]].
  
\bibitem{Bertacca:2014wga} 
  D.~Bertacca, R.~Maartens and C.~Clarkson,
  JCAP {\bf 1411}, 013 (2014)
  [arXiv:1406.0319 [astro-ph.CO]].
  
\bibitem{Yoo:2014sfa} 
  J.~Yoo and M.~Zaldarriaga,
  Phys.\ Rev.\ D {\bf 90},  023513 (2014)
  [arXiv:1406.4140 [astro-ph.CO]].
  
\bibitem{DiDio:2014lka} 
  E.~Di Dio, R.~Durrer, G.~Marozzi and F.~Montanari,
  JCAP {\bf 1412}, 017 (2014)
  [JCAP {\bf 1506}, no. 06, E01 (2015)]
  [arXiv:1407.0376 [astro-ph.CO]].

\bibitem{Bertacca:2014hwa} 
  D.~Bertacca,
  arXiv:1409.2024 [astro-ph.CO].

\bibitem{Umeh:2012pn} 
  O.~Umeh, C.~Clarkson and R.~Maartens,
  Class.\  Quantum Grav.\  {\bf 31}, 202001 (2014)
  [arXiv:1207.2109 [astro-ph.CO]].

\bibitem{Umeh:2014ana} 
  O.~Umeh, C.~Clarkson and R.~Maartens,
  Class.\ Quant.\ Grav.\  {\bf 31}, 205001 (2014)
  [arXiv:1402.1933 [astro-ph.CO]].

\bibitem{Kaiser:1987qv} 
  N.~Kaiser,
  Mon.\ Not.\ Roy.\ Astron.\ Soc.\  {\bf 227}, 1 (1987).

\bibitem{Verde:1998zr} 
  L.~Verde, A.~F.~Heavens, S.~Matarrese and L.~Moscardini,
  Mon.\ Not.\ Roy.\ Astron.\ Soc.\  {\bf 300}, 747 (1998)
  [astro-ph/9806028].


\bibitem{BeltranJimenez:2010bb} 
  J.~Beltran Jimenez and R.~Durrer,
  Phys.\ Rev.\ D {\bf 83}, 103509 (2011)
  [arXiv:1006.2343 [astro-ph.CO]].
  

\bibitem{Jeong:2008rj} 
  D.~Jeong and E.~Komatsu,
  Astrophys.\ J.\  {\bf 691}, 569 (2009)
  [arXiv:0805.2632 [astro-ph]].

  
\bibitem{Carlson:2009it} 
  J.~Carlson, M.~White and N.~Padmanabhan,
  Phys.\ Rev.\ D {\bf 80}, 043531 (2009)
  [arXiv:0905.0479 [astro-ph.CO]].


\bibitem{Hagstotz:2014qea} 
  S.~Hagstotz, B.~M.~Schaefer and P.~M.~Merkel,
  arXiv:1410.8452 [astro-ph.CO].
  
\bibitem{Bernardeau:2001qr} 
  F.~Bernardeau, S.~Colombi, E.~Gaztanaga and R.~Scoccimarro,
  Phys.\ Rept.\  {\bf 367}, 1 (2002)
  [astro-ph/0112551].

\bibitem{Sheth:1999mn} 
  R.~K.~Sheth and G.~Tormen,
  Mon.\ Not.\ Roy.\ Astron.\ Soc.\  {\bf 308}, 119 (1999)
  [astro-ph/9901122].

\bibitem{Sheth:2001dp} 
  R.~K.~Sheth and G.~Tormen,
  Mon.\ Not.\ Roy.\ Astron.\ Soc.\  {\bf 329}, 61 (2002)
  [astro-ph/0105113].


\bibitem{Switzer:2013ewa} 
  E.~R.~Switzer {\it et al.},
Mon.\ Not.\ Roy.\ Astron.\ Soc.\  {\bf 434}, L46 (2013)  
  [arXiv:1304.3712 [astro-ph.CO]].
\end{thebibliography}



\end{document}